%% file: 16455.tex
\documentclass[twocolumn,traditabstract,longauth]{aa}

\input{16455_a.tex}

\usepackage{graphicx,amsmath}
\usepackage{epsf}
\usepackage{epsfig}
\usepackage{natbib}
\usepackage{rotating}
\usepackage{url}
\usepackage{txfonts}
\newcommand{\changes}{}

\begin{document}
\input{16455_b.tex}
  \title{\textit{Planck} Early Results 21: Properties of the interstellar medium in the Galactic Plane}
\date{ }

\authorrunning{The \Planck\ collaboration}
\titlerunning{Properties of the interstellar medium in the Galactic plane}
\abstract{
\Planck\ has observed the entire sky from 
 30 GHz to  \getsymbol{HFI:center:frequency:857GHz:units}. 
The observed foreground emission contains contributions from different phases of the interstellar 
medium (ISM). 
We have separated the observed Galactic 
emission into the different gaseous components (atomic, molecular and ionised) in each of a number 
of Galactocentric rings. This technique provides the necessary information to study dust properties 
(emissivity, temperature, etc.), as well as other emission mechanisms as a function of Galactic radius.
Templates are created for various Galactocentric radii using velocity information from atomic (neutral hydrogen) and molecular ($^{12}$CO) 
observations.
The ionised template is assumed to be traced by free-free emission as observed by {\it WMAP}, while  408 MHz emission is used to trace the 
 synchrotron component.
 Gas emission not traced by the above templates, namely ``dark gas'', as evidenced using \Planck\ data,
is included as an additional template, the first time such a component has been used in this way.
These templates are then correlated with each of the \Planck\ 
frequency bands, as well as with higher frequency data from {\it IRAS} and DIRBE along with radio data at 1.4\,GHz. The emission per 
column density of the gas templates allows us to create distinct spectral energy distributions (SEDs) 
per Galactocentric ring and in each of the gaseous tracers from 1.4\,GHz to 25 THz ($12\micron$).
The resulting SEDs allow us to explore the contribution of various emission mechanisms to the \Planck\ signal. Apart from the thermal 
dust and free-free emission, we have probed the Galaxy for anomalous (e.g., spinning) dust as well as synchrotron emission.
We find the dust opacity in the solar neighbourhood, $\tau/N_{\rm H} = 0.92\pm0.05\times10^{-25} {\rm cm}^2$ {\changes at 250 $\mu$m}, with no significant variation 
with Galactic radius, even though the dust temperature is seen to vary from over 25 K to under 14 K. 
Furthermore, we show that anomalous dust emission is present in the atomic, molecular and dark gas phases throughout the Galactic disk. 
Anomalous emission is not clearly detected
in the ionised phase, as free-free emission is seen to dominate. {\changes The derived dust propeties associated with the dark gas phase 
are derived but
do not allow us to reveal the nature of this phase.}
  For all environments, the  anomalous emission is consistent with rotation from polycyclic aromatic hydrocarbons (PAHs) and, 
according to our simple model, accounts 
for $(25\pm5)\%$ (statistical) of the total emission at 
30 GHz. 
}


\maketitle
\section{Introduction}\label{sec:intro}

In order to understand our own Galaxy, it is necessary to explore  
Galactic Plane  ($|b|\la10\degr$), 
where we are able to observe the emission coming from a large range of distances. 
However the observed emission is the sum of a large number of line-of-sight components, often probing very different environments. 

Several previous studies \citep{Bloemen1986, Bloemen1990, Giard1994, Sodroski1997, Paladini2007} have separated observed integrated emission into 
a number of Galactocentric radii in order to study its properties as a function of Galactic position and in different phases of the interstellar gas 
(e.g., atomic, molecular and ionised). The radial velocity of the gas is used to separate the Galactic gas emission into a number of Galactocentric rings 
and then  the spectral energy distribution of each ring/gas phase is fitted with a physical model of dust and gas emissions.
These methods have been used successfully in previous studies to map out basic properties of the ISM throughout the Galaxy.

For instance, \cite{Giard1994} demonstrated that polycyclic aromatic hydrocarbons (PAHs) are a ubiquitous component
 of the interstellar medium, and \cite{Bloemen1990} 
 showed that the dust temperature 
decreases with  distance to the Galactic Centre in a way which is fully consistent with an
 exponential decrease of the interstellar radiation field (ISRF) in the stellar disk. 
\cite{Sodroski1997} suggested that the abundance of large dust grains
within each gas phase exhibits a gradient that is equivalent, 
within the uncertainties, to the metallicity gradient in the 
Galactic disk. \cite{Paladini2007} showed that the dust in molecular clouds appears to be heated in a significant way 
by young massive stars still embedded in their parent clouds.

With the advent of the \Planck\ satellite\footnote{\Planck\ (http://www.esa.int/Planck) is a project of the European Space Agency (ESA) with instruments provided by two scientific consortia funded by ESA member states (in particular the lead countries: France and Italy) with contributions from NASA (USA), and telescope reflectors provided in a collaboration between ESA and a scientific consortium led and funded by Denmark.}, it is now possible to perform an inversion on the emission arising from the entire 
infrared, millimetre and centimetre range, 
making it possible to determine the 
dust and gas properties in many different environments in the Milky Way.
The High Frequency Instrument (HFI) channels (100--857 GHz) allow us to properly constrain the big grain temperature and emissivity, while using the Low Frequency Instrument 
(LFI) channels (30--70 GHz), it is possible to constrain non-thermal emission mechanisms
such as free-free, synchrotron or anomalous dust. We have used these data along with other ancillary data to perform 
a large-scale, low-resolution analysis of the Milky Way ISM emission.

Knowledge of the dust emission and physical conditions throughout the Milky Way provides a self-consistent context for other \Planck\ studies: for example, 
the environmental effect on the properties of cold cores \citep{planck2011-7.7a,planck2011-7.7b} can be evaluated, 
the modelling of individual anomalous dust regions can be placed in a Galaxy-wide context \citep{planck2011-7.2}, 
or the Milky Way values can simply be compared to other galaxies \citep{planck2011-6.4a,planck2011-6.4b}.

In Sect.~\ref{sec:data} we describe the data used in the study, including both those to be inverted and those used as templates to represent a 
particular phase 
of the ISM. In Sect.~\ref{sec:method}, we describe how we optimise the separation of the gas observation into 
a number of Galactocentric rings, as well as how we report uncertainties on our best solution. We present the results in Sect.~\ref{sec:results}, 
and discuss potential sources of uncertainty and bias in Sect.~\ref{sec:discussion}. Finally we conclude in Sect.~\ref{sec:conclusion}.

\section{Data}\label{sec:data}

There are two categories of data used in this study, the data to be inverted, and data used to create templates which 
will be used to perform the inversion.

\subsection{Data to be inverted}

We use data from $12\micron$ (25 THz) to 1.4\,GHz in order to study the dust and gas properties in a self consistent manner
 throughout the Milky Way. Here we briefly present these data in order of decreasing (increasing) frequency 
(wavelength).

The Infrared Astronomical Satellite (\textit{IRAS}), a joint project of the US, UK, and the Netherlands \citep{Neugebauer1984}, performed a survey of 98$\%$ of
the sky at four wavelengths: 12, 25, 60, and $100\micron$.
We use a reprocessed version of this data set \citep[IRIS,][]{MivilleDeschenes2005} 
which benefits from a better zodiacal light subtraction and from a calibration and zero level compatible with {DIRBE}, as well as from a better destriping.

The DIRBE instrument \citep[Diffuse Infrared Background Experiment,][]{Hauser1998} 
 on board the \textit{COBE} satellite imaged the full
sky in 10 broad photometric bands from 1 to $240\micron$ with a beam of 0.7\degr. The bands at 140 and $240\micron$ allow us to bridge the gap between \textit{IRAS}
 and the highest frequencies of HFI, thus covering the peak of the thermal dust spectral energy distribution (SED).

The  \Planck\ data we used for this analysis are the DR2 LFI and HFI maps, with the cosmic microwave background (CMB) removed, as  
described in detail in \cite{planck2011-1.6}  and \cite{planck2011-1.7}, respectively. 
It consists of a set of 9 frequency maps with central frequencies of 
 \getsymbol{LFI:center:frequency:30GHz},
 \getsymbol{LFI:center:frequency:44GHz},
 \getsymbol{LFI:center:frequency:70GHz},
  \getsymbol{HFI:center:frequency:100GHz},
 \getsymbol{HFI:center:frequency:143GHz},
 \getsymbol{HFI:center:frequency:217GHz},
 \getsymbol{HFI:center:frequency:353GHz},
  \getsymbol{HFI:center:frequency:545GHz} and
 \getsymbol{HFI:center:frequency:857GHz:units}), 
smoothed to a common resolution of 1\degr FWHM, at a common HEALPix \citep{Gorski2005} pixelization resolution of $N_{\rm side}$=256 ($\sim$15' pixels). 
For simplicity, in the remainder of the text we will refer to the first three bands as 30, 44 and 70 GHz, respectively.
The CMB component that was removed from these maps was found by the Needlet Internal Linear Combination (NILC) CMB extraction method presented in \cite{planck2011-1.7}. 

Finally, to provide constraints at radio frequencies we use
the full-sky 1.4\,GHz data from \citep{Reich1982, Reich1986, Reich2001}.

All of the above data sets have  been  smoothed with a Gaussian function 
to a common resolution of 1\degr FWHM and gridded to a common HEALPix $N_{\rm side}$ of 256.

\subsection{Templates}

\begin{figure*}
  \centering
  \includegraphics[width=0.9\linewidth]{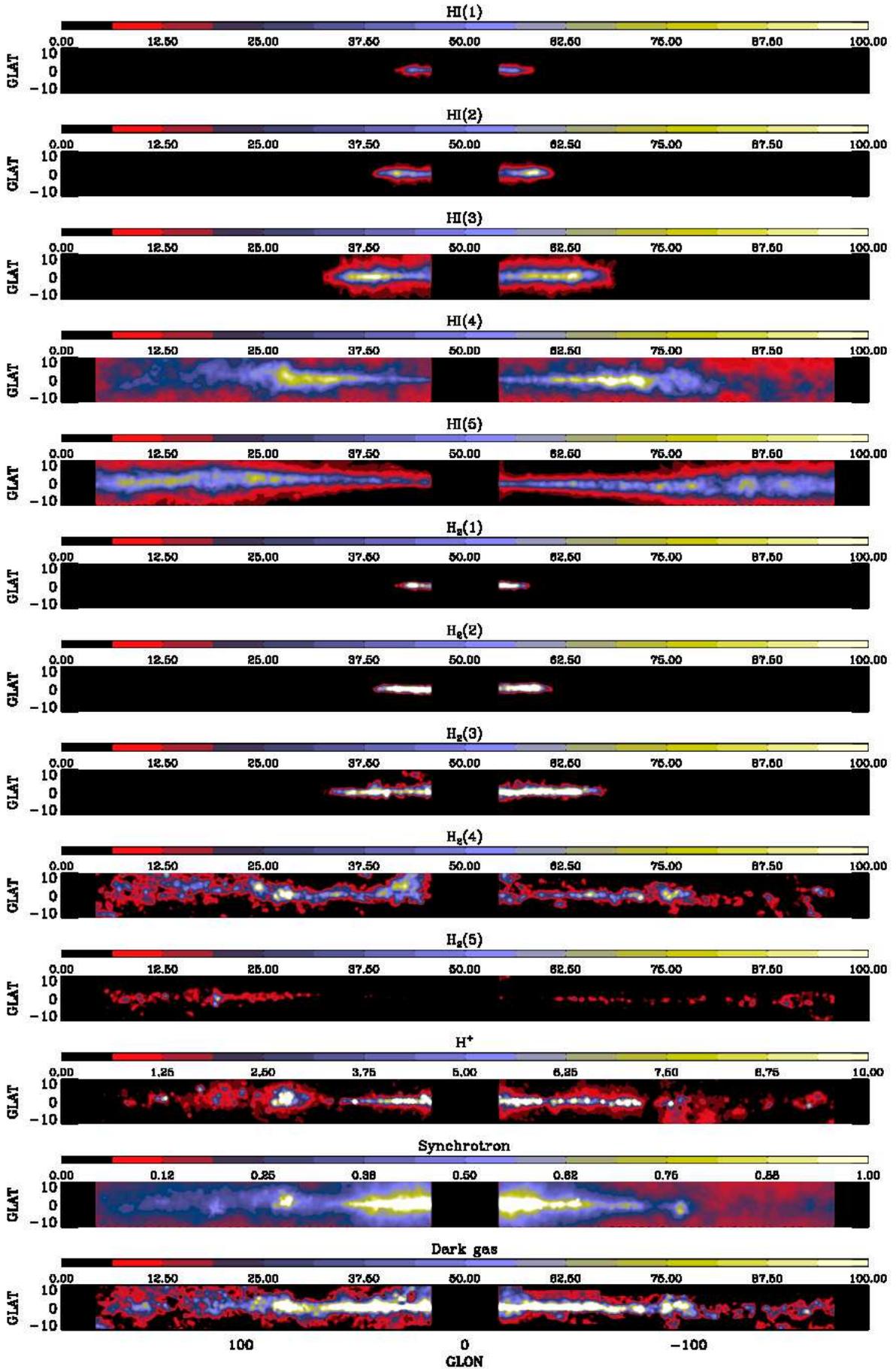}

  \caption{Templates used in the inversion. The top five images are the \ion{H}{i} templates, the following five are the molecular templates, all of which are presented 
in increasing Galactocentric radius {\changes (the radii defining each ring are described in Sect. \ref{sec:optimrings})}.
The final three templates are the free-free, synchrotron, and dark gas templates, respectively. The units are all expressed in 
$10^{20}$ atoms, except for synchrotron, expressed in MJy\,sr$^{-1}$. The centre and anti-centre regions have been masked as no distance information can be obtained 
from the radial velocity of the gas.}
  \label{fig:templates}
\end{figure*}
\label{sec:templates}

\subsubsection{Atomic phase}
We aim to characterise the different emission mechanisms
 in the different phases of the interstellar medium (atomic, molecular and ionised). For the atomic phase we use the 
Leiden/Argentine/Bonn Galactic \ion{H}{i} Survey \citep{Kalberla2005}.
The survey results are expressed in antenna temperature, $T_{\rm A}$. 
In order to convert this into column density, the effect of optical depth should be taken into account, for example
\citep{Binney1998}
\begin{equation}
\label{eq:nh}
  N_{\rm \ion{H}{i}} = 1.82 \times 10^{-18} \times \int_{v_1}^{v_2}  T_s\times\ln\left(1-\frac{T_A}{T_s}\right)^{-1}dv \, {\rm cm}^{-2},
\end{equation} 
for a given spin temperature ($T_{\rm s}$). 
Recent studies in the plane of the Milky Way \citep{Dickey2003,Dickey2009} advocate the use of a higher spin temperature of around 250 K. This higher value 
is justified on the grounds that along any line of sight there is a mix of cold and warm \ion{H}{i} gas, and so an effective 
spin temperature is needed to correct for the opacity. This will 
most certainly vary as a function of Galactic longitude. However, constraining this variation is beyond the scope of this study. We 
have therefore assumed a constant hydrogen spin temperature of 250 K for the entire Galactic Plane. 
Varying this value in the range 150-400 K has only a small effect on our results and they remain consistant within the quoted uncertainties.

\subsubsection{Molecular phase}
We assume that the bulk of the molecular gas mass is at first order well traced by CO emission \citep{Tielens2005}.
The CO data we use come from the Composite CO Survey of \cite{Dame2001}. 
The column density of molecular hydrogen can be expressed as a function of antenna temperature ($T_\mathrm{CO}$) 
\begin{equation}
  N_{\rm H_2} = 2 \times X_{\rm CO} \times \int_{v_1}^{v_2}  T_{\rm CO} \, dv.
\end{equation}
where $X_{\rm CO}=N({\rm H_2})/W_{\rm CO}$ is the ratio of molecular hydrogen column density to the velocity-integrated intensity of the $^{12}$CO line.
 We use a value for $X_\mathrm{CO}$ which is compatible with the recent values from the \textit{FERMI} collaboration \citep{Abdo2010,Ackermann2010},  $X_{\rm CO} = 1.8\times 10^{20}$, which results from a 
study on diffuse gamma-ray emission in the Galactic plane.
 
Both the atomic and molecular data contain line of sight velocity information. It is therefore possible to use this information to 
separate the emission into a set of Galactocentric rings, as a given line of sight velocity measurement uniquely identifies a Galactocentric distance, assuming circular orbits. 
This should not be confused with heliocentric distances for which there is a distance ambiguity corresponding to the near or far side of the derived Galactocentric orbit. 
The velocity information alone does not allow one to 
determine if the observed gas is on the near side of the determined orbit or on the far side. 
We separate the observed \ion{H}{i} and CO observations each into a set of 5 Galactocentric rings. 
The reasons for choosing this decomposition is explained in Sect.~\ref{sec:optimrings}

\subsubsection{Dark gas}
\begin{figure}
  \centering
  \includegraphics[width=\linewidth]{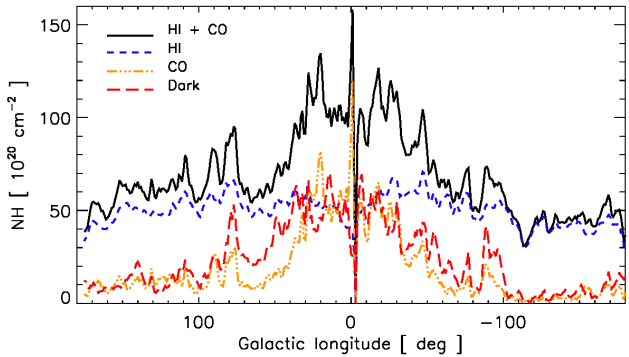}
 \caption{Longitude profile of \ion{H}{i}, CO and dark gas. {\changes The black solid line shows the total contribution from \ion{H}{i} \& CO}. The dark gas represents a significant fraction of the gas column density, 
and dominates the CO outside of the molecular ring.}
  \label{fig:dark_gas_lon_prof}
\end{figure}

It has been shown \citep[e.g.,][]{Reach1994,Grenier2005} that the CO  and \ion{H}{i} do not account for all the atomic and molecular gas in interstellar clouds: there is in addition a dark gas phase. This phase is also extensively studied in \cite{planck2011-7.0}, where the total gas column density, as traced by \ion{H}{i} and CO observations, 
is correlated with the optical depth of the dust calculated using {\it IRAS} $100\micron$ along with \Planck\ HFI data at 350 and $500\micron$ (857
 and 545 GHz).
The correlation is linear at low and high column densities, but it departs from linearity at column densities of $7.4\times10^{20}<N_{\rm H}<5.0\times10^{21}$.
This excess is attributed to thermal emission by dust associated with a dark gas phase, undetected by \ion{H}{i} and CO measurements. 

{\changes 
The dark gas column density used here is defined in \cite{planck2011-7.0} as
\begin{equation}
N_{\rm H}^{X} = (\tau_{\rm D}-\tau_{\rm M})/\left(\frac{\tau_{\rm D}}{N_{\rm H}}\right)^{\rm ref},
\end{equation}
where $\tau_{\rm D}$ is the thermal dust optical depth, $\tau_{\rm M}=({\tau_{\rm D}}/{N_{\rm H}})^{\rm ref} [N_{\ion{H}{i} + 2X_{\rm CO}W_{\rm CO}}])$ 
is the modelled dust opacity and $({\tau_{\rm D}}/{N_{\rm H}})^{\rm ref}$ is the reference dust emissivity measured in low $N_{\rm H}$ regions.}
We use their dark gas estimate at  857 GHz  as an additional template.

The dark gas phase represents 
a significant fraction of the total gas column density, as shown in Fig.~\ref{fig:dark_gas_lon_prof}. The \ion{H}{i} dominates over almost the entire longitude range, and the 
CO is strongest in the molecular ring $40\degr>l>-40\degr$. Outside of the molecular ring the dark gas is stronger than the CO, except towards the anti-centre 
direction where the dark gas and CO are roughly comparable. The proportion of dark gas to total hydrogen column density ranges from roughly $10\%$ in the anti-centre direction up 
to nearly $60\%$ towards the Galactic centre.

\subsubsection{Synchrotron emission}

An additional emission mechanism which is important in the frequency range we are studying and is not traced by usual gas tracers is synchrotron emission,
which arises from the acceleration of cosmic-ray electrons in magnetic fields. This emission follows approximately a power law but the spectral index most certainly 
varies on the sky and as a function of frequency. More than 90\% of the observed synchrotron emission arises from a diffuse component \citep{Bennett2003} 
and in the Galactic plane recent studies have shown that there is no hardening of the synchrotron spectrum \citep{Finkbeiner2004a, Boughn2007}, but 
that emission from spinning dust is likely to dominate between 8 and 60\,GHz. 
 For our synchrotron template we have chosen
the  408\,MHz all-sky survey from \cite{Haslam1982}, which traces the diffuse, soft synchrotron component.

\subsubsection{Ionised phase}

To represent the ionised component we have used the Maximum Entropy Method (MEM) free-free template from the {\it WMAP} seven year data \citep{Jarosik2010}.
The intensity map ($I_{\nu}$) has been 
 converted to ionised hydrogen column density as described by \cite{Sodroski1997}:
\begin{equation}
  N_{\rm H_{II}} = 1.2 \times 10^{15}\, T_{\rm e}^{0.35}\, \nu^{0.1}\, n_{\rm eff}^{-1}\, I_{\nu}
\end{equation}
for effective electron density $n_{\rm eff}$ and electron temperature $T_{\rm e}$.
Following \cite{Sodroski1989}, we use  $n_{\rm eff}=10$ cm$^{-3}$ and  $T_{\rm e}=8000$ K.
In the future, the availability of radio recombination line (RRL) surveys will provide a more direct measurement of the free-free component in the Galaxy
and its distribution separated into distinct Galactocentric bins using the radial velocity information included in such surveys.

 As has been done for the data, the templates have all been smoothed to 1\degr\, FWHM 
 and gridded to an $N_{\rm side}$ of 256. These templates are all displayed in Fig.~\ref{fig:templates}.

\section{The Galactic inversion}\label{sec:method}
\begin{figure*}
  \centering
  \epsfig{figure=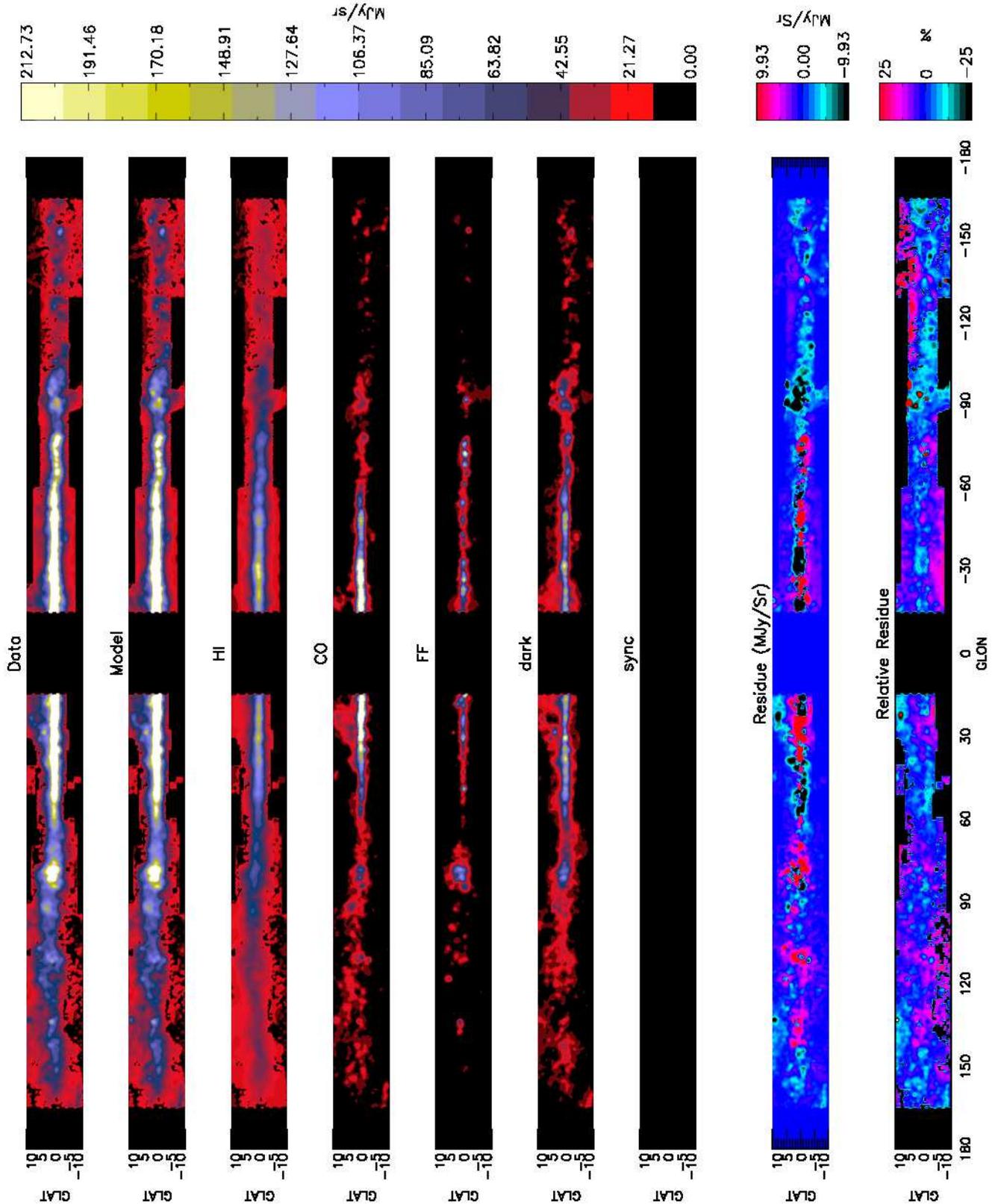, angle=90}
  \caption{Results of the inversion for the 857 GHz band. From  top to bottom are: observed emission; modelled emission; atomic contribution; molecular contribution; 
ionised contribution; dark gas contribution; synchrotron contribution; residual; and relative residual. All images except for the residual maps are at 
the same intensity scale. The centre and anti-centre regions have been masked, as the kinematic distance method is inapplicable to these regions. 
Areas where no CO observations are available have also been masked.}
  \label{fig:857sol}
\end{figure*}

\begin{figure*}[ht!]
  \centering
  \epsfig{figure=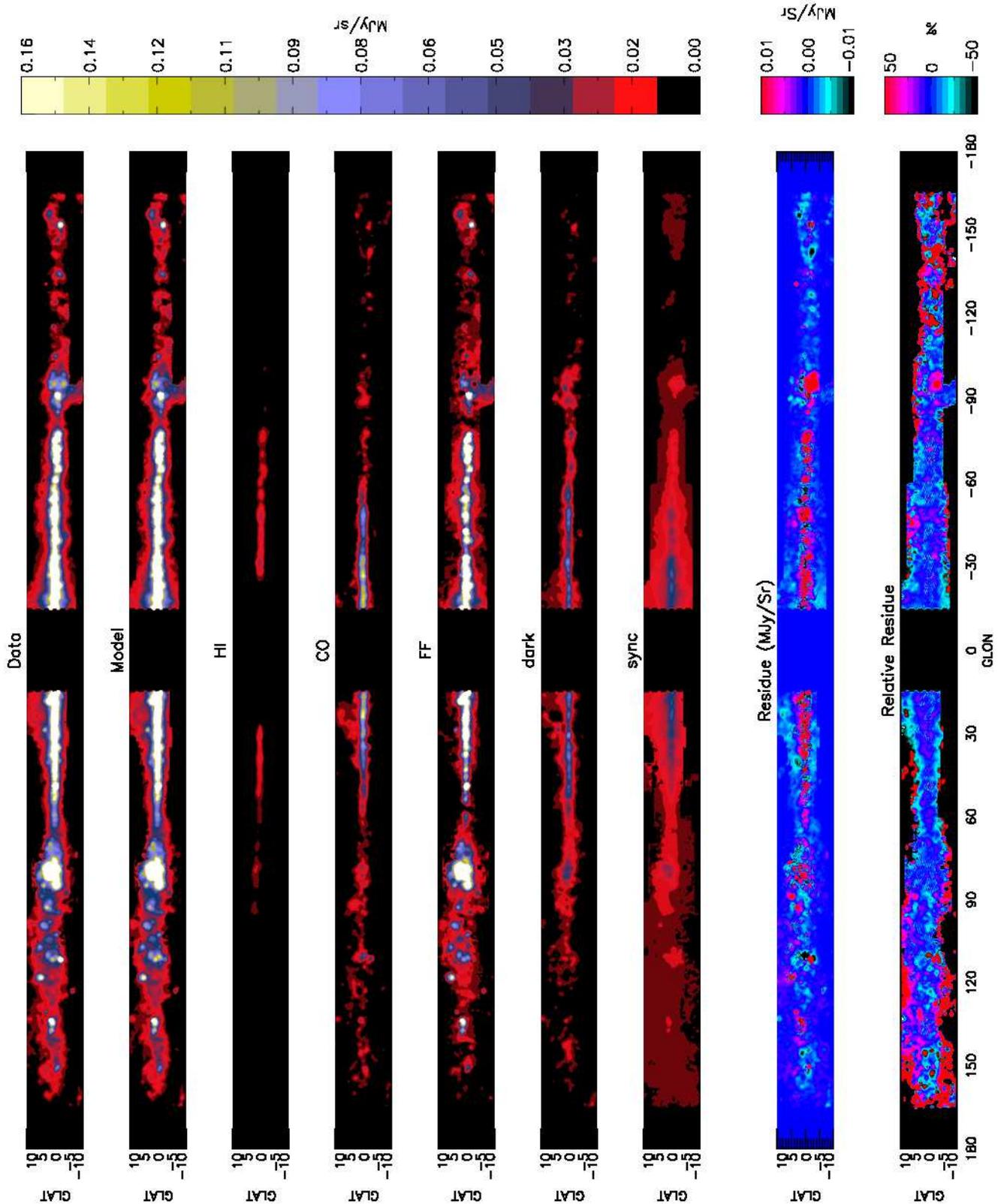, angle=90}
   \caption{Same as Fig.~\ref{fig:857sol} for the 30 GHz band.}
  \label{fig:30sol}
\end{figure*}
\begin{figure*}[ht!]
  \centering
  \epsfig{figure=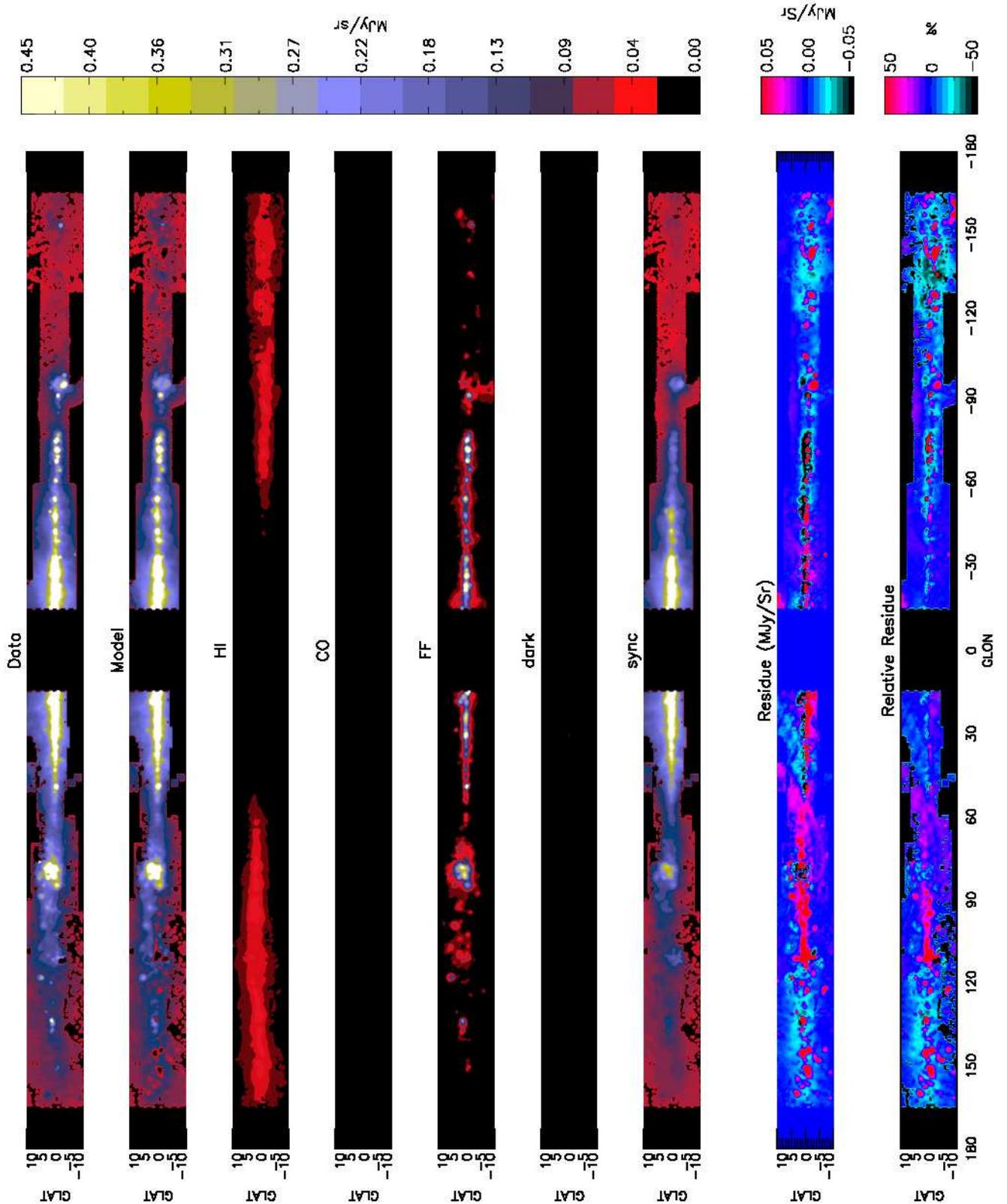, angle=90}
  \caption{Same as Fig.~\ref{fig:857sol} for the 1.4\,GHz band.}
  \label{fig:1.4sol}
\end{figure*}
Component separation techniques have been widely tested in the framework of multi-frequency observations of the CMB, especially in the context of \Planck\ 
\citep[see][, for a review of the methods developed within the \Planck\ collaboration]{Leach2008}. These methods exploit spectral and spatial 
correlations and independence in observations made at separate frequencies, as well as external constraints and physical modelling, 
in order to distinguish between different physical sources of emission. Most of them are designed to efficiently separate CMB from 
foregrounds and possibly foregrounds into ``components'': (Galactic synchrotron,
free-free and dust emissions; extra-galactic and far-IR point sources; Sunyaev-Zeldovich effect, etc.

The approach presented here is different from these methods as it tries to quantify the separate components responsible for diffuse Galactic emission.
 Nevertheless it is not strictly speaking a component separation method, in the sense presented above, as it fits 
one map of the sky with spatial templates at different Galactocentric distances
 and is not exploiting the spectral signature of each component to perform the separation.

\subsection{Inversion method}

We assume that each of the frequency maps ($I_{\nu}$) can be expressed as a linear combination of the spatial templates described in Sect.~\ref{sec:templates} 
and presented in Fig.~\ref{fig:templates}. 
Then, for each frequency map, we are solving for the mean emissivity of each template (in units of MJy sr$^{-1}$ / $10^{20} {\rm atoms}$) which best reproduces the 
observed data. Note that we are not taking into account any spatial variations of the emissivity within a given template, but are assuming a constant value 
for each of our templates.
The modelled emission can be written, for any frequency map $I_{\nu}(x,y)$, as:
\begin{eqnarray}
  \label{eqn:inversion}
  I_{\nu}^{m}(x,y) &=& \Sigma^{n}_{i=1} \left( \epsilon^i_{\rm H_I}(\nu)N_{\rm H_I}^i(x,y) +\epsilon^i_{\rm H_2}(\nu)N_{\rm H_2}^i(x,y) \right ) \nonumber \\
 &+& \epsilon_{\rm H_{II}}(\nu)N_{\rm H_{II}}(x,y) + \epsilon_{\rm s}(\nu)N_{\rm s}(x,y) \nonumber \\
 &+& \epsilon_{\rm d}(\nu)N_{\rm d}(x,y) + C_{\nu},
\end{eqnarray}
where the sum is over a number ($n$) of Galactocentric rings ($i$) for the atomic and molecular phases, denoted $N_{\rm H_I}$ and  $N_{\rm H_2}$ respectively.
$N_{\rm H_{II}}, N_{\rm s}$ and $N_{\rm d}$ denote the ionised, synchrotron and dark gas components, respectively, and $C_{\nu}$ is a constant. Note that the last three templates 
do not have any rings, as we have no way of separating the information into different Galactocentric distances and so they are a function of frequency only.
 As such, the quoted $\epsilon$ are a function of $\nu$ only.

In order to find the values for the different emissivities,  $\epsilon$, 
we minimise the chi-squared value defined by
\begin{equation}
\label{eqn:chi2}
  \chi^2 = \Sigma_{j=1}^{N_{\rm pixels}}\frac{(I_{\rm mod}(j)_{\nu} - I(j)_{\nu})^2}{\sigma_{\nu}^2},
\end{equation}
where $\sigma_{\nu}$ is the noise in the map $I_{\nu}$, estimated as the median value of the intensity for $|b|>70\degr$.

In our case, the difference between model and observations is not at all dominated 
by the low noise present in the \Planck\ bands, but by the simplistic model that we are adopting for the inversion. As such we suppose that the noise
does not vary spatially.

The solution which minimises Eq.~\ref{eqn:chi2} is found via matrix inversion. By writing the problem as 
\begin{equation}
b_i = a_j \cdot A_{ij}, 
\end{equation}
where $b$ is the observed intensity map, $A$ is an $N\times M$ matrix of the different templates and $a$ is the vector of the parameters. The solution is 
simply \citep{Press1992}
\begin{equation}
\label{eqn:ata}
a= (A^TA)^{-1} \times A^Tb\;.
\end{equation}

The optimal mathematical solution which solves  Eqs.~\ref{eqn:inversion}--\ref{eqn:chi2} may not be physical; in particular, it may include 
negative emissivities. However, we do not have any cases where the emissivity is more than one sigma below zero, so remaining consistent with zero. 
As such we do not place any constraints on the parameters of the inversion.

\subsection{Galactocentric ring optimisation}
\label{sec:optimrings}
The observed radial velocities of \ion{H}{i} and CO are used to separate the observed emission into a number of 
Galactocentric rings, assuming circular orbits for the gas and using the rotation curve from \cite{Clemens1985}.
The choice of the distance intervals for the rings is not arbitrary, but is subject to a number of constraints.
Choosing too many would create a set of rings that would be highly correlated and result in a degenerate solution.
An optimised set of rings has been defined by minimising the correlation between rings, and thus
minimising the degeneracy between the various components of the inversion.
The total number of rings and their thickness are also driven by the requirement of a minimal number of data falling into
each ring in order to constrain it and separate its contribution from the others. Further, we have chosen to fix the solar circle as 
one of the rings, including all velocity channels that contain high latitude, low velocity clouds.

The ring optimisation process consists first in dividing the \ion{H}{i} and CO templates
into a large number of small rings, and then building the covariance matrix ($A^TA$, see Eq.~\ref{eqn:ata}) associated
with this highly discrete basis of annuli. An iterative process is then performed:
neighbouring rings with the highest correlation are merged and the new covariance matrix
is updated with a reduced number of annuli. The process continues, merging the rings which have 
strongest correlation between them, until a reasonable number of rings is obtained.
This leads to the following decomposition: $R_{\rm GAL} =  [1.5, 4.2, 5.4, 7.1, 9.8, 25.0]$\,kpc, where the values define the ring edges.

\subsection{Estimating uncertainty}
\label{sec:jack}
The result of the inversion is 
an average gas emissivity at each frequency, in each gas phase and in each ring (when available).
However, within a given gas phase the dust and gas properties may 
vary significantly and we wish to report this uncertainty. 
In other words, the residual differences between the model and the data are not due to 
noise in the data, meaning that the uncertainty on the resulting emissivities cannot be 
computed analytically (i.e., the residuals are not well-modelled by the uniform Gaussian noise implicit in the solution, Eq.~\ref{eqn:ata}).  
The residuals are rather due to unknown differences between the templates and the data.  
An alternate method is then needed to estimate the uncertainties on the results.

The dust properties in the northern Galaxy may differ from those in the southern part, and of course the variation may be even more localised than that.  
To quantify this uncertainty 
we perform jackknife tests by masking approximately $30\%$ of the Galaxy in 30 random blocks, each 3\degr wide, and performing 
the inversion on the unmasked zones. This is repeated 500 times, and the rms value of the jackknife tests is used 
to provide 
the total variation of the emissivity within each ring and gas phase.
We therefore provide robust estimates of the dust emission and provide their variation for each ring and gas phase.

It should be stressed that this estimation does not take into account the errors on either the \ion{H}{i} opacity correction 
nor the uncertainty on the assumed $X_{\rm CO}$ value for each phase globally.

\section{Properties of the ISM in the Galactic plane}\label{sec:results}

The above inversion method has been applied to 16 different frequency bands from $12\micron$ (25\,THz) to 1.4\,GHz.
The results are presented in table form in the appendix (Table~\ref{tab:emm_res}).

Figures~\ref{fig:857sol}, \ref{fig:30sol} and \ref{fig:1.4sol} compare our best-fit model to the data for three frequencies 
(two \Planck\ frequencies, 857  and 30 GHz, as well as the  1.4\,GHz radio frequency) 
and the contributions of the different gas components to the model.
The model reproduces the data rather well, especially when one notices the scale of the  residual. 
The contours in the relative residual are drawn every 25\%, so nearly the full zone used in the inversion has residuals below this level.
 
The main contributors to the 857 GHz map (see Fig.~\ref{fig:857sol})
are the atomic and molecular phases
 (including the dark gas) with a modest contribution from the ionised component.
At  30 GHz (Fig.~\ref{fig:30sol})  the emission is a combination of the molecular, ionised, dark gas and synchrotron, with very little hydrogen.
As for the solution at 1.4\,GHz (Fig.~\ref{fig:1.4sol})
it is the  synchrotron component that dominates the solution, but with a significant contribution from the ionised component.

\subsection{Dust in thermal equilibrium}

\begin{figure}[t!]
  \centering
  \includegraphics[width=0.85\linewidth]{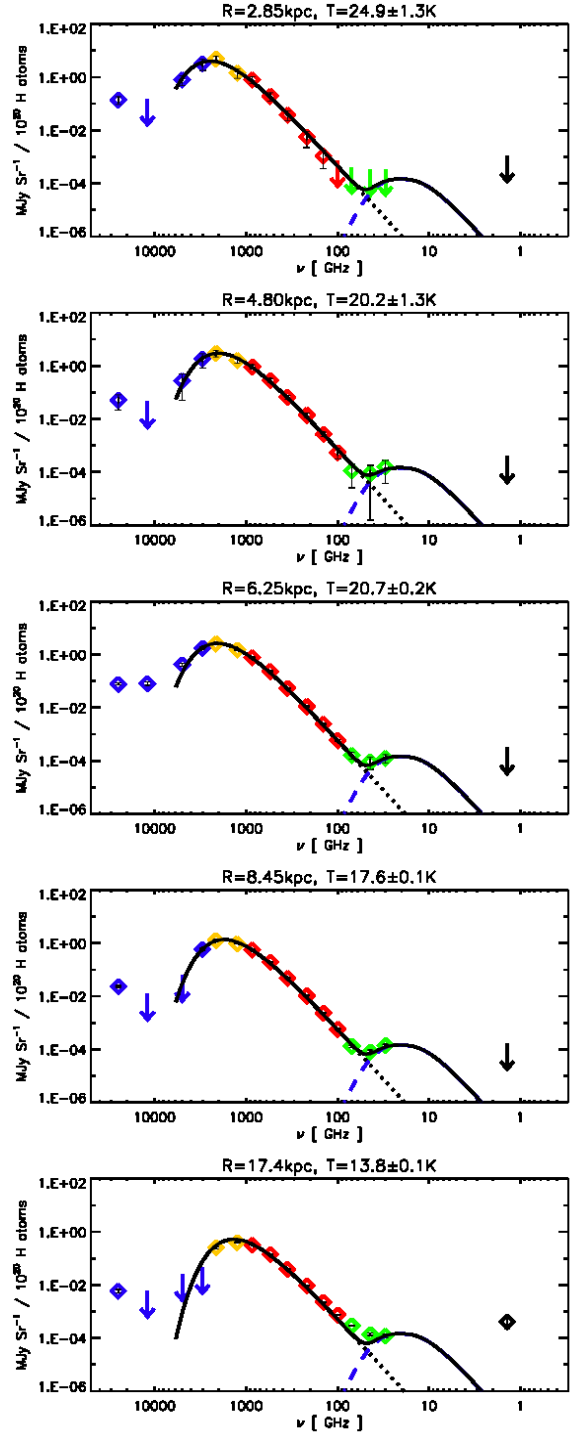}
  \caption{SEDs for the atomic component, each sorted 
in increasing Galactocentric radius. The colour of the symbols refer to the mission the data came from: {\it IRAS} (blue),  DIRBE (yellow), HFI (red), LFI (green) and 1.4\,GHz (black). The SEDs also show 
various fitted laws: the dotted line is the thermal dust SED, the short dashed line spinning dust in an atomic medium,
and the solid line the sum of all contributions. The spinning dust contribution is not a fit, but 
simply the result of using typical values in the model (see Sect.~\ref{sec:ame} for details). 
The contribution of the atomic component in the outer Galaxy to the 1.4\,GHz signal is lower than the noise level derived for the 1.4\,GHz map and so should not be viewed as significant.
}
  \label{fig:HI_SED}
\end{figure}

\begin{figure}[t!]
  \centering
  \includegraphics[width=0.85\linewidth]{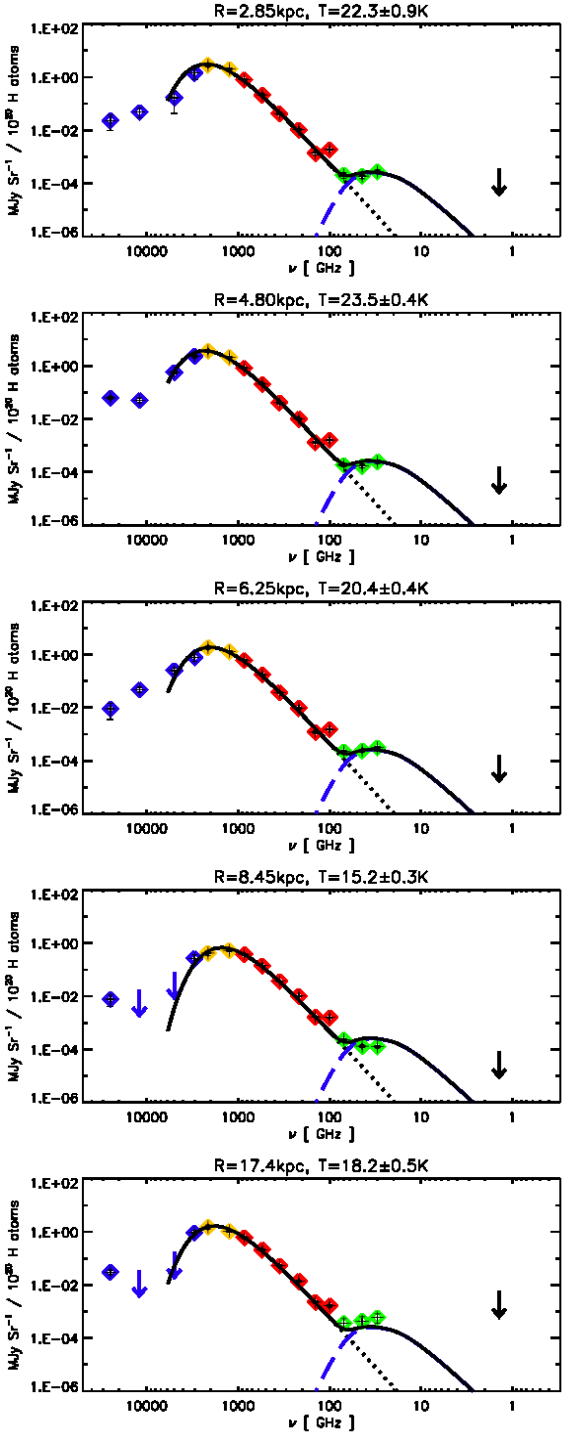}
  \caption{SEDs for the CO component, each sorted 
in increasing Galactocentric radius. The different lines and colours are as described in Fig.~\ref{fig:HI_SED}, except for the long dashed line which represents spinning dust in a molecular 
medium.}
  \label{fig:CO_SED}
\end{figure}

We have constructed SEDs for each of our templates across all frequencies that we have analysed. 
Where the emissivity is compatible with zero, we plot the $2\sigma$ upper limit as a downward arrow.

\begin{figure}
  \centering
  \includegraphics[width=0.8\linewidth]{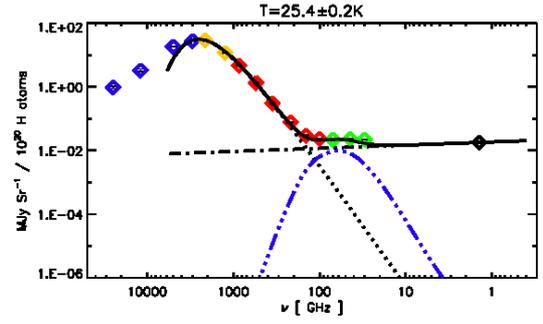}
  \caption{The resulting SEDs for the free-free component. The different lines and colours are as described in Fig.~\ref{fig:HI_SED},
 except for the dash-dot line and the dash triple dot line which represent free-free emission and  spinning dust in an ionised medium, respectively.}
  \label{fig:FF_SED}
\end{figure}

\begin{figure}
  \centering
  \includegraphics[width=0.9\linewidth]{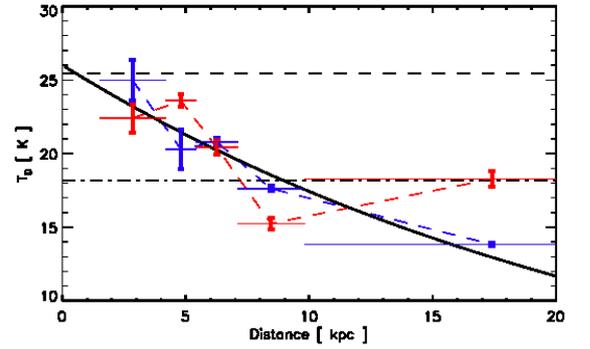}
  \caption{Average dust temperature as a function of Galactocentric radius for the atomic (blue) and molecular (red) components. 
{\changes The dust temperature in the ionised phase and the dark gas phase are shown by the  dashed line and dot-dashed line, respectively.} 
The solid line shows the behaviour of the ISRF with Galactic radius.
Horizontal bars show the range in distance over which the mean temperature applies.
 The temperature of dust mixed with atomic \ion{H}{i} 
 decreases with radius, consistent with the fact that this component is heated by the ambient ISRF.
 Dust in the molecular component is heated by embedded star formation. 
}
  \label{fig:temp_dist}
\end{figure}

We have fit the SEDs using a sum of thermal dust (big grains), spinning PAH molecules, free-free and synchrotron emission.
The large dust grains in thermal equilibrium are well described by a modified blackbody of the form
\begin{equation}
  I_{\nu} \propto B_{\nu} \frac{\nu^{\beta}}{\nu_0} = \frac{2h\nu^3}{c^2(e^{h\nu/kT} -1)}\frac{\nu^{\beta}}{\nu_0}\;,
\end{equation}
where $T$ is the dust temperature and $\beta$ is the emissivity spectral index. 
When fitting using both $T$ and $\beta$ as free parameters, we find a mean value of $\beta=1.8\pm0.1$ throughout the Galaxy. 
However, to increase the reliability of the fit, we follow \cite{planck2011-7.12,planck2011-7.13,planck2011-7.0} and fix $\beta=1.8$. 
We use all data points down to 143\,GHz in the fit, but exclude 217\,GHz in the molecular and dark gas phases due to contamination from CO line emission.
 All points are colour corrected assuming $\nu I_{\nu}={\rm constant}$ 
as described in \cite{planck2011-1.7}. The SEDs for the atomic, molecular and ionised phases are shown in Figs.~\ref{fig:HI_SED}, \ref{fig:CO_SED} and \ref{fig:FF_SED}, 
respectively.
 We are  able to constrain the dust temperature 
for each component and in the various Galactocentric radii. The resulting temperature profile is shown in Fig \ref{fig:temp_dist}. The 
temperature in the atomic phase is seen to decline steadily with increasing Galactocentric distance. This decrease in temperature
 follows the trend of the ambient ISRF \citep{Mathis1983}, shown by the solid line in  Fig \ref{fig:temp_dist}, 
indicating that the dust in the atomic phase is predominantly heated by it. 

 The dust in the molecular component, however, is seen to be more steady,  peaking in the molecular ring ($3 \la R_{\rm G} \la 6$ kpc) and once again 
outside the solar circle. This phase seems to be heated by the presence of star formation. The ionised phase (dotted line) is seen to be dominated by grains 
that are on average  much warmer than 
either of the other two phases. These results are in agreement with previous Galactic analyses \citep{Sodroski1997,Paladini2007}.
The temperature within the solar circle ($T=17.6\pm0.1\,{\rm K}$) is completely in agreement with the average value at high Galactic latitude \citep{planck2011-7.12}.

\begin{figure}
  \centering
  \includegraphics[width=\linewidth]{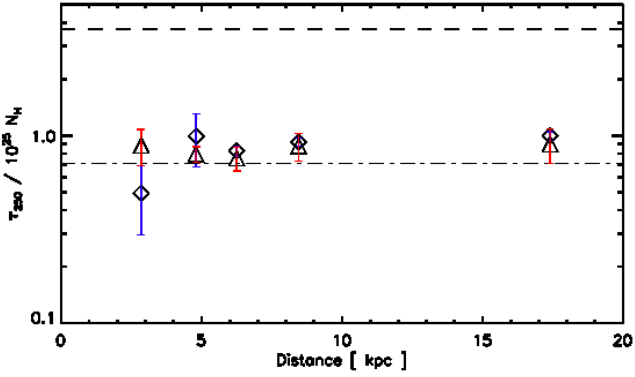}
  \caption{Dust opacity at $250\micron$ as a function of Galactocentric radii for \ion{H}{i} (blue, diamonds) and CO (red, triangles). 
The dashed line  and the dash dot line show the dust opacity for the ionised phase and the dark gas phase, respectively.}
  \label{fig:emm_res}
\end{figure}

The dust opacity at a given wavelength is defined as 
 $\tau_{\lambda} / N_{\rm H}$ 
where $\tau_{\lambda} =\epsilon_{\lambda} / B(\nu,T)$ is the dust optical depth.
The dust opacity in the atomic phase for the solar circle is 
equal to $0.92\pm0.05\times10^{-25}\,{\rm cm}^2$ {at 250 $\mu$m}, slightly below, but in agreement with, the same value found by \cite{planck2011-7.12} at 
high Galactic latitude. 
The variation of the opacity as a function of Galactic radius is shown in Fig.~\ref{fig:emm_res}.
We do not detect any significant variation of the dust opacity as a function of Galactic radius, in either the atomic or molecular phases.
Note, however, that 
the opacity of the molecular component is completely degenerate with the chosen value for $X_{\rm CO}$. By increasing  $X_{\rm CO}$, we decrease 
the opacity. Further, the  appropriate spin temperature for the \ion{H}{i} opacity correction may differ from the constant 250\,K we have assumed.

\begin{figure}
  \centering
  \includegraphics[width=0.8\linewidth]{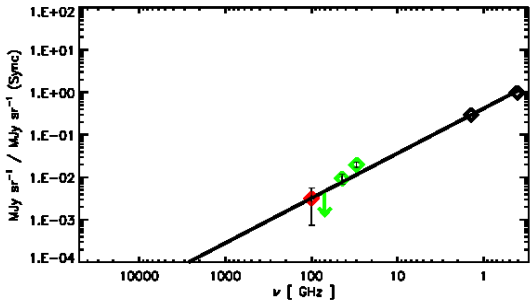}
  \caption{The resulting SEDs for synchrotron component. The different lines and colours are as described in Fig.~\ref{fig:HI_SED}. The point at 408\,MHz is not part of the inversion, but 
has been plotted here to show consistency.}
  \label{fig:SYNC_SED}
\end{figure}

Finally, after fitting the thermal dust in the molecular phase, there is an obvious excess at 100 and 217\,GHz due to the presence of 
CO line emission in these \Planck\ bands \citep{planck2011-1.7}. This is a clear indication that the inversion is working as it should, as this excess
 is present in the molecular phase.
The excess with respect to the fit is $16.2\pm 1.0$ and $20.2 \pm 4.5\muK_{\rm CMB}/{\rm K}_{\rm RJ}\kms$ at 100 and 217\,GHz,  respectively. 
 \cite{planck2011-1.7} find $14.2\pm1.0$ and $44.2\pm6.0$  at 100 and 217\,GHz, respectively. At 100\,GHz our value is slightly higher, which 
is to be expected as it is simply an excess with respect to $^{12}$CO and does not take into account the other CO isotopes present along 
the line of sight. The difference in the values at 217\,GHz is not as easy to understand, but may come from our simple assumption of a fixed $\beta$.

\subsection{Analysis of non-thermal SED components}

The SEDs for all components at frequencies lower than 100\,GHz  show non-thermal dust emission features. The emission mechanisms at work 
in this spectral range are thermal bremsstrahlung (free-free), synchrotron emission, as well as a third commonly referred to as 
anomalous dust emission.

\subsubsection{Synchrotron}

The SED for the synchrotron component resulting from our inversion is shown in Fig.~\ref{fig:SYNC_SED}. 
 A power-law has been recovered in the two lowest 
frequency LFI points and the 1.4\,GHz radio point.

For electrons with a power-law distribution of energies $N(E) \propto E^{-p}$
the frequency dependence of the emission is characterised
by antenna temperature $T(\nu)\propto\nu^{\,\beta}$ with spectral index $\beta = -(p +3)/2$, with typically $\beta\sim -3$. 
As we are not working in antenna temperature, but flux density,  the power law takes  the form $I_{\nu}\propto \nu^{\,\alpha}$ with $\alpha = \beta-2$.

We have performed a fit of the rising synchrotron spectrum with a power law and obtained a spectral index $\alpha= -1.0\pm0.1$. 
This value 
is slightly flatter than the average value found between 408\,MHz and 23\,GHz by \textit{WMAP} 
\citep{Dunkley2009, Davies2006}. 
However, this value falls within the range of values found in the Galaxy \citep[$-0.5<\alpha<-1.1$]{Bennett2003}, and most likely 
represents the stronger power injection in the Galactic plane.
There is no evidence in the data for any hardening of the synchrotron spectrum 
below 70\,GHz.

\begin{figure*}
  \centering
  \includegraphics[width=\linewidth]{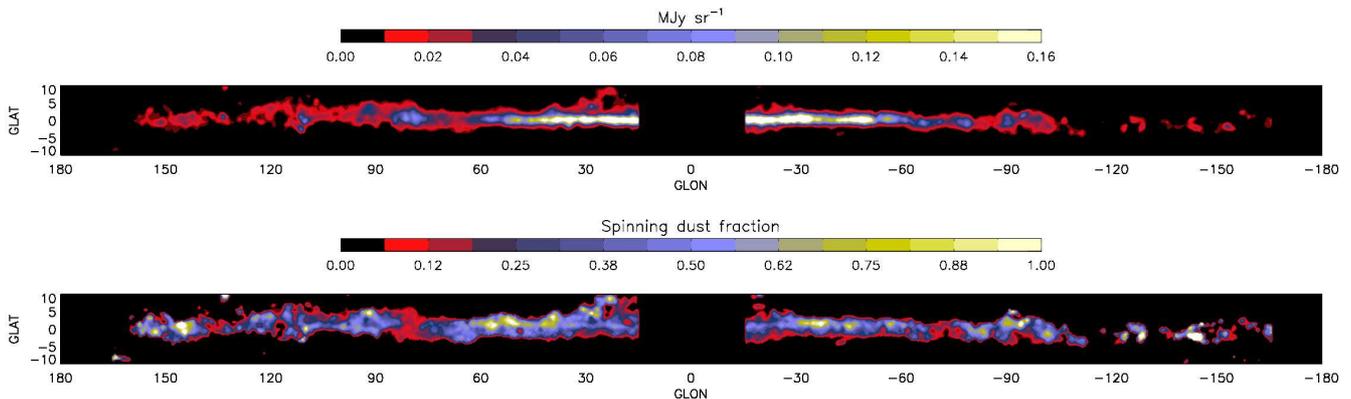}
  \caption{Depiction of the spinning dust emission at 30 GHz in the Galactic plane  in MJy sr$^{-1}$ (top) and as a fraction of the total30 GHz  signal (bottom). 
    In making this image, it is assumed that the 30 GHz emission associated with the atomic, molecular and dark gas phases are solely due to spinning dust. Some features 
may be simply due to residuals in our inversion; however, this illustrative view highlights regions where spinning dust may be particularly strong compared to other emission 
mechanisms.}
  \label{fig:SPIN_MAP}
\end{figure*}

\subsubsection{Free-free}
\label{sec:freefree}

Free-free emission  dominates the ionised hydrogen template (Fig \ref{fig:FF_SED}) at frequencies below 143\,GHz. This emission follows 
a power-law with a spectral index of $\alpha=-0.1$ for $70$\,GHz$<\nu<1.4$\,GHz. 
As we have used the {\it WMAP} free-free template
to trace the ionised medium,  it is interesting to compare the total free-free emission that we attribute to this phase with 
the {\it WMAP} estimate. 
 Using the power-law 
fit shown in Fig.~\ref{fig:FF_SED}, we can extrapolate to the frequency of our choice and,
 using our H+ template,  produce an estimate of the free-free emission in any of the {\it WMAP} bands. 

Performing this at 23\,GHz, we recover $\sim80\%$  of the emission that the \textit{WMAP} team find in their MEM free-free component separation. 
Using radio recombination lines (RRLs) from the \ion{H}{i} Parkes Zone of Avoidance (ZOA) survey,  \cite{Alves2010} recently showed that 
  diffuse ionised emission can be recovered in the Galactic plane using RRL surveys. They then compared the level of free-free derived with 
the  {\it WMAP} MEM  estimate. As in this work, they showed it to be too high
in the Galactic plane by at least $10\%$, albeit  in a much smaller ($8\degr \times 8 \degr$) region.  

\begin{figure}
  \centering
  \includegraphics[width=0.8\linewidth]{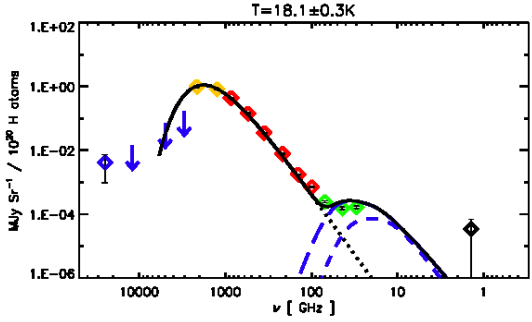}
  \caption{The resulting SED for the  dark gas component. 
The different lines and colours are as described in Fig.~\ref{fig:HI_SED}. The dashed and short-dashed lines shows the spinning dust contribution from a molecular phase, and from an atomic phase, respectively.
}
  \label{fig:DARK_SED}
\end{figure}

Looking at the molecular SEDs (Fig.~\ref{fig:CO_SED}), and the dark gas SED (Fig.~\ref{fig:DARK_SED}), the LFI points depart from the thermal dust
and seem to describe a free-free spectrum. It is not possible, however, to describe the three LFI points {\it and} the radio point using a free-free power law, except 
for the first and last molecular templates. Even so, all the radio points in the molecular templates are upper limits and  
the predicted free-free spectrum would be above the 4-$\sigma$ level for three of the templates, showing that the data are not compatible with this model.
In addition,  if this were indeed free-free emission, 
our estimate of the total Galactic free-free emission would surpass that of the {\it WMAP} MEM free-free estimate by over 30\%. This seems unlikely in light of 
the \cite{Alves2010} finding discussed above.
In that case, the  LFI-radio points in the atomic, molecular and the dark gas phases are 
anomalous microwave emission.
Indeed, \cite{Boughn2007} recently showed that the dominant emission mechanism at 20\,GHz is spinning dust, and so is identified as anomalous microwave emission. 

\subsubsection{Anomalous microwave emission}
\label{sec:ame}

 Anomalous microwave emission, arising between 10 and 100\,GHz and first detected in the 1990s \citep{Kogut1996}, cannot be explained
 using classical emission mechanisms known in this frequency range \citep{Banday2003}. Indeed it is too bright to be free-free emission, and it is 
not polarised as thermal dust or synchrotron emissions would be. However, it is correlated with dust IR emission and especially 
with the interstellar PAH and very-small-grain (VSG) emission in the mid-IR \citep{Casassus2006, Ysard2010b, Scaife2010}. 
Anomalous emission was first proposed by \cite{Draine1998} to come from rapidly rotating PAHs \citep{Rouan1992}, the spinning dust grains. 
 Since, it has been observed in various interstellar environments: 
dark clouds \citep{Watson2005,Casassus2006, Scaife2009, Dickinson2010}; \ion{H}{ii} regions \citep{Dickinson2007, Scaife2008}; planetary nebulae \citep{Casassus2007} 
and diffuse interstellar medium \citep{MivilleDeschenes2008}. 
Several processes can excite or damp the grains' rotation: photon emission (IR and radio); gas/grain interactions; 
formation of H$_2$ molecules on the grain surface; or photoelectric emission.

Here we investigate the anomalous component in CO-correlated emission. 
All the rings exhibit a flattening of the SED in the three LFI bands (Fig. \ref{fig:CO_SED}), 
as well as upper limits on the  emission at 1.4\,GHz.
This clear excess in the microwave range cannot be free-free emission due to the reasons discussed in Sect.~\ref{sec:freefree}.
The peak frequency of this excess lies approximately 
from 20 to 60\,GHz.

We attempt to fit the excess with the spinning dust model described in \cite{Silsbee2010}. Models show 
that spinning dust emission is sensitive to the grain size distribution, to the intensity 
of the ISRF,
and to the density of the medium in 
which the grains are embedded \citep{Ali-Haimoud2009, Ysard2010a}. The smallest grains, namely the PAHs, are responsible for most of the spinning dust emissivity. 
Fitting the thermal dust emission with a modified blackbody, we estimate the radiation field to be 
close to the standard ISRF \citep{Mathis1983, Porter2008}. 
To describe the small dust grains, we use two grain populations: PAHs and VSGs. We adopt a size distribution that is 
the sum of two log-normal components \citep{Weingartner2001}. 
The first one is peaked at 3\,nm (VSG), and the second one at 0.6\,nm (PAHs), in line with \cite{Compiegne2011}, with the width of both log-normal components set to 0.4. 
Most of the spinning dust emission comes from the smallest grains, the second log-normal component, for which we assume an abundance equal to the value measured in the solar neighbourhood.
Finally we describe the grains' environment as molecular gas
with typical parameters ($n_H = 350$ cm$^{-3}$ and $T_{\rm gas} = 20$ K). 

The fit to the data with this simple model is surprisingly good, and shows that not only is anomalous emission present in the molecular phase throughout the 
Milky Way, but that spinning dust provides a compelling model with which to describe it. The predominance of spinning dust in the molecular 
phase is in agreement with more detailed modelling of individual anomalous emission regions \citep{planck2011-7.2}.

The ISRF could in theory also have an effect on this emission mechanism, 
and the big grain temperature gives us an idea of its fluctuations:
$\chi = (T / 17.5{\rm K})^{\beta+4}$, where $\chi$ is dimensionless and equal to one in the solar neighbourhood. 
However this range of ISRF fluctuations 
has little effect on the spinning dust spectrum  \citep{Ali-Haimoud2009, Ysard2010a}, so we assume a solar value for all SEDs.
The average density of the molecular gas may vary with radius, which would have enough impact 
to significantly alter the fit. However, we have not proceeded with a detailed fit of this component, as 
there are many parameters and the solution would be degenerate (e.g. grain size distribution, gas density, ISRF intensity, as well as radiative transfer considerations). 
We have kept the analysis simple, which allows us to say that we have 
definitely detected anomalous microwave emission, that it is associated with the molecular phase, and that spinning dust provides 
a good model with which to understand the spectrum of this excess emission. Some of this emission could be explained by free-free emission, but it cannot be the dominant 
emission mechanism in this phase.

To study the possibility that spinning dust emission is present in either the atomic or ionised phase, 
we have also estimated its contribution to these. The different environments will have an impact on the spinning dust emission, so 
we have adapted the gas parameters in the modelling.
For the atomic phase we have used the results of \cite{Heiles2003} and modelled 
the atomic ISM as composed of 40\% cold neutral medium (CNM) and 60\% warm neutral medium (WNM).
The parameters for these phases are $n_{\rm H} = 30$\,cm$^{-3}$ and $T_{\rm gas} = 100$\,K for the CNM, and  $n_{\rm H} = 0.4$\,cm$^{-3}$ and $T_{\rm gas} = 6000$\,K for the WNM.
As can be seen in Fig.~\ref{fig:HI_SED}, the SEDs show that spinning dust emission is able to reproduce the departure from the thermal dust 
spectrum for $\nu<70$\,GHz. The 1.4\,GHz point in the last atomic template is almost certainly an artifact. The atomic and synchrotron templates 
both describe diffuse smooth emission in the Galaxy, leading to some ``crosstalk'' between the two components. However it should be noted that  when the emissivity found for the atomic phase  
in the outer Galaxy is multiplied by the column density of hydrogen in the template, the total emission is very low, lower than the noise level derived for the 1.4\,GHz map. As such, it should not be viewed as significant.

For the ionised phase, 
we use the same gas parameters as in Sect.~\ref{sec:templates}, namely $n_{\rm H} = 10$\,cm$^{-3}$ and $T_{\rm gas} = 8000$\,K.
The SED in Fig.~\ref{fig:FF_SED} shows that spinning dust may also be present in this phase, but 
that it is completely dominated by free-free emission. In fact, the situation may be even worse, as we have assumed solar value PAH abundance, 
whereas recently \cite{Dobler2009} showed that the PAH abundance in this type of region is about a factor of three lower.

Assuming that the atomic, molecular and dark gas contributions at 30 GHz are due solely to spinning dust, and that the ionised phase has a negligible contribution, we have produced a Galactic-plane image of spinning dust emission. 
This provides a complementary image to that presented in  \cite{planck2011-7.2}, as it has been obtained using a different approach.
The intensity map is shown at the top of Fig.~\ref{fig:SPIN_MAP}; at the bottom we show the fraction of the total signal at 30 GHz  due to spinning dust. Several 
bright regions show up in this map, highlighting areas where spinning dust emission is the dominant source of emission at these frequencies.
Overall,  spinning dust accounts for $25\pm5\%$(statistical) of the total emission at 30 GHz. This error estimate does not take into account any systematic uncertainty 
due to our simple model, which is difficult to quantify, and so will likely be higher.

\subsection{Dark gas}

\begin{figure*}[t!]
  \centering
 \includegraphics[width=0.7\linewidth]{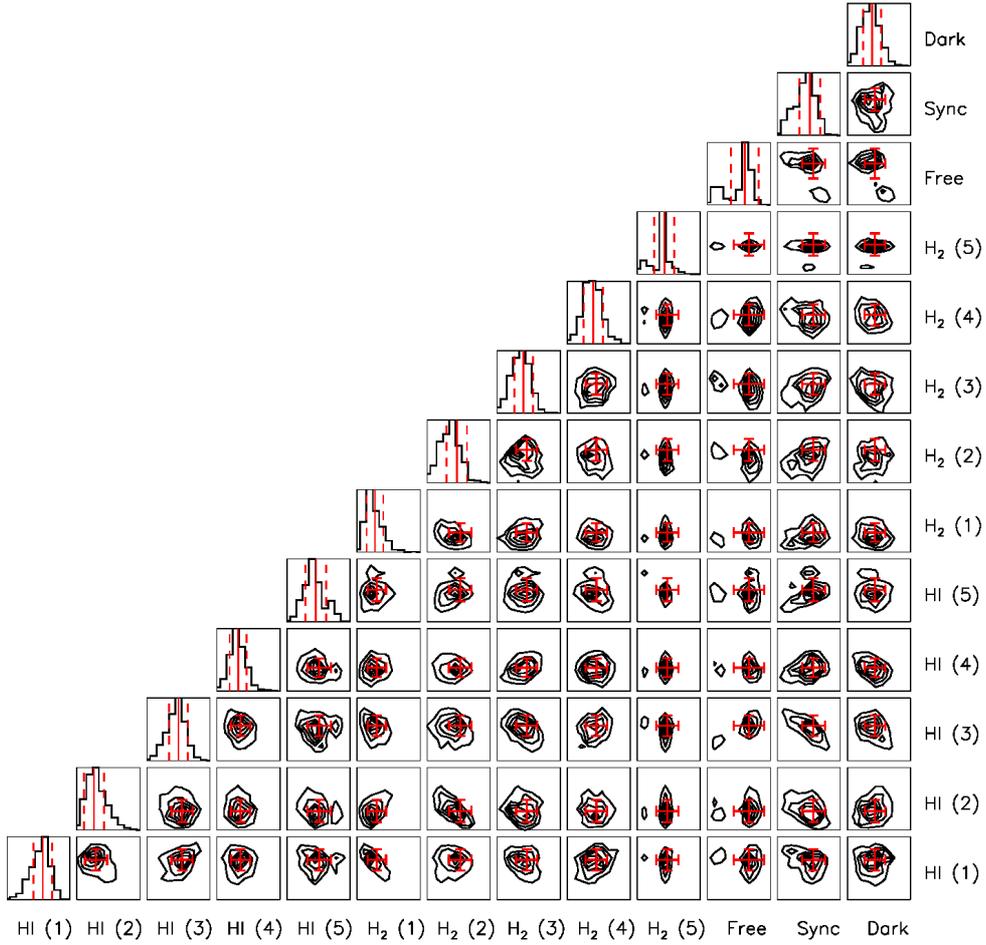}  
\caption{Correlation matrix of the emissivities for each ring at
30 GHz. Overplotted in red are the results with corresponding error bars. One-dimensional histograms replace the contour plots 
where only one component is concerned.}
  \label{fig:cross_correl_30}
\end{figure*}
Dark gas has been detected in numerous studies \citep{Reach1994, Grenier2005} as an infrared, and a correlated infrared/gamma-ray, excess over and above what is expected given the gas column density 
derived from \ion{H}{i} and CO data.
The inversion performed without this component results in positive and negative residuals, as the other components compete to account 
for this missing gas phase. Not only does adding this component lower our reduced $\chi^2$ fits {\it across all bands}, but it also allows us to 
examine the nature of this dark component.

Looking at Fig\ref{fig:DARK_SED}, the resulting SED for this component looks quite similar to both the atomic and  molecular SEDs. Its temperature does not differentiate it immediately from either 
of the two phases, and it 
shows a low-frequency tail beyond the thermal dust SED. These points look similar to the 
anomalous dust in the molecular phase (dashed line); however, the spinning dust model lies above the LFI points. Assuming that the spinning dust is in an atomic phase (dotted line), it can be 
seen that it does not predict the observed emission. It thus seems to be more of a mix of the two phases, with a larger influence from the molecular phase.
The different parameters discussed in 
Sect.~\ref{sec:ame} could also be altered, but we do not have enough data to constrain such a fit.

The interpretation of these results is somewhat complicated for several reasons. Firstly, the amount of dark gas is not expected to be independent of 
Galactic position, but using a 2D template we are forced to assume just that. Secondly,
 the estimate of this gas phase was obtained at high galactic latitude and 
we are using an extrapolation into the Galactic plane. 
It would be preferable to use an independent template, derived from gamma-ray observations for example, in future inversion analyses.
Reconstructing the line-of-sight distribution of the dark gas would enable a more exact determination of the properties of the dark gas, by coupling the analysis with  a
3D extinction map \citep[e.g.][]{Drimmel2003,Marshall2006,Sale2009}.
This type of analysis will be considered in a future study of the Galactic inversion. 

In summary, the dark gas SED is consistent with  molecular gas, most likely diffuse H$_2$ and optically thick CO. The best fit 
 at longer wavelengths comes from the spinning dust model, in a molecular medium.

\section{Possible sources of bias}
 Our inversion algorithm determines the flux density per column density of each template, and hence
 the result depends on the different templates themselves. Care should therefore be
taken to use the most physically representative input for each phase of the ISM that we 
want to include in the inversion. 
Nonetheless, the jackknife tests described in Sect.~\ref{sec:jack} show that the method is self-consistent.
Moreover, many different combinations for the \ion{H}{i}+CO Galactocentric rings has been attempted, and 
the main results presented here remain coherent with those. 
However, the results are notably sensitive to one particular input, the synchrotron template. 
Without it, the \ion{H}{i} templates 
trace the Galactic synchrotron emission, as the \ion{H}{i} observations trace the large-scale smooth structure 
of the Milky Way, similar to the diffuse soft synchrotron emission.
 This artificial component in the \ion{H}{i} templates then masks any contribution for 
the anomalous microwave emission. We have attempted to trace the synchrotron emission with 
two different synchrotron templates: the WMAP MCMC synchrotron template \citep{Gold2009}; and the 408\,MHz 
full sky survey \citep{Haslam1982}. We have decided to use the latter as it more successfully removes 
the synchrotron contribution from the \ion{H}{i} templates and allows us to 
detect anomalous emission in the atomic phase.

\label{sec:discussion}
\subsection{Correlation between parameters and uniqueness of the solution}

In Fig.~\ref{fig:cross_correl_30} we show the correlation between the different emissivities of our best solution at 30 GHz, where 
many emission mechanisms are present and the inversion is more challenging.
These were obtained from the result of the jackknife tests presented in Sect.~\ref{sec:jack}.
It can be seen that for almost all the templates the correlation is low. Three elements are to be highlighted. 
The first one is that the emissivities for the outer Galaxy molecular template, as well as the ionised template, both have pronounced departures from a Gaussian
 distribution. 
Looking at Fig.~\ref{fig:templates}, both templates are patchy; the masking of several regions simultaneously will change the morphology of each map significantly and therefore 
 have a large impact 
on the derived emissivity.

Secondly, rings 1--3 of the atomic and molecular phase are slightly anti-correlated (the contours are elliptical). This region of the 
Galaxy is indeed difficult to disentangle, and due to the low filling factor of these rings the algorithm has less to work with than in 
the solar circle, for example. Nevertheless the error bars deduced from the jackknife analysis account for this slight degeneracy.
Higher resolution maps will help reduce this crosstalk between templates in future analyses.

Thirdly a small bias can be seen, as the results of the inversion are not always centred in the middle of the jackknife 
distributions. We calculated this bias by looking at the difference between  the 
mean deviation of each result with respect to the jackknife realisations, in units of result uncertainty. That is the 
bias $B$ per template $i$ and per  band $\nu$ can be written:
\begin{equation}
  B(i,\nu) = \frac{1}{\sigma(i,\nu)}  \frac{1}{N_{\rm r}}  \Sigma_{r=1}^{N_{\rm r}} \left[\epsilon(i,\nu)-\epsilon_J(r,i,\nu) \right],
\end{equation} 
where $N_{\rm r}$ is the number of jackknife realisations,  $\epsilon(i,\nu)$ and ${\sigma(i,\nu)}$ are the result and uncertainty 
of the inversion, and $-\epsilon_J(r,i,\nu) $ is the result of one jackknife realisation.
We have found this effect to be small, with a median bias of $0.005 \sigma$ over all templates and frequencies,  and with 
a mean absolute deviation of $ 0.08 \sigma$. In all cases, the bias is always lower than $0.25 \sigma$.
\begin{figure}
  \centering
 \includegraphics[width=0.9\linewidth]{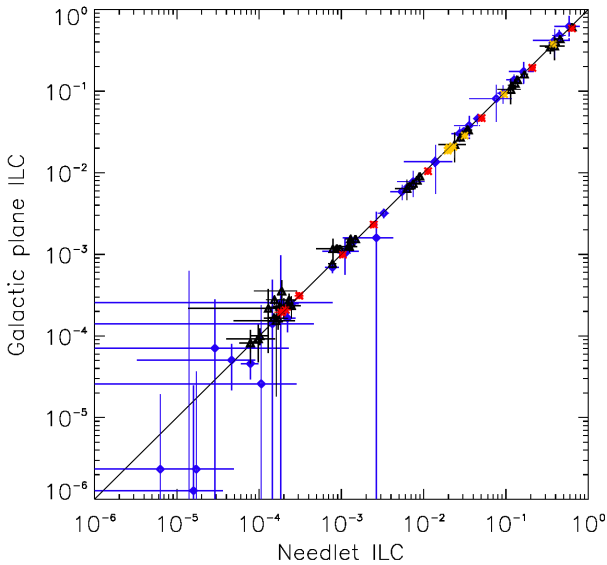}  
\caption{Comparison of emissivities derived after application of two different CMB subtractions, the official Needlet ILC and 
the Galactic plane ILC. The atomic templates are the blue points, molecular are black, ionised is orange, dark gas is red 
and synchrotron is green.}
  \label{fig:emm_JA_cmb}
\end{figure}
\begin{figure*}[ht!]
  \centering
 \includegraphics[width=0.9\linewidth]{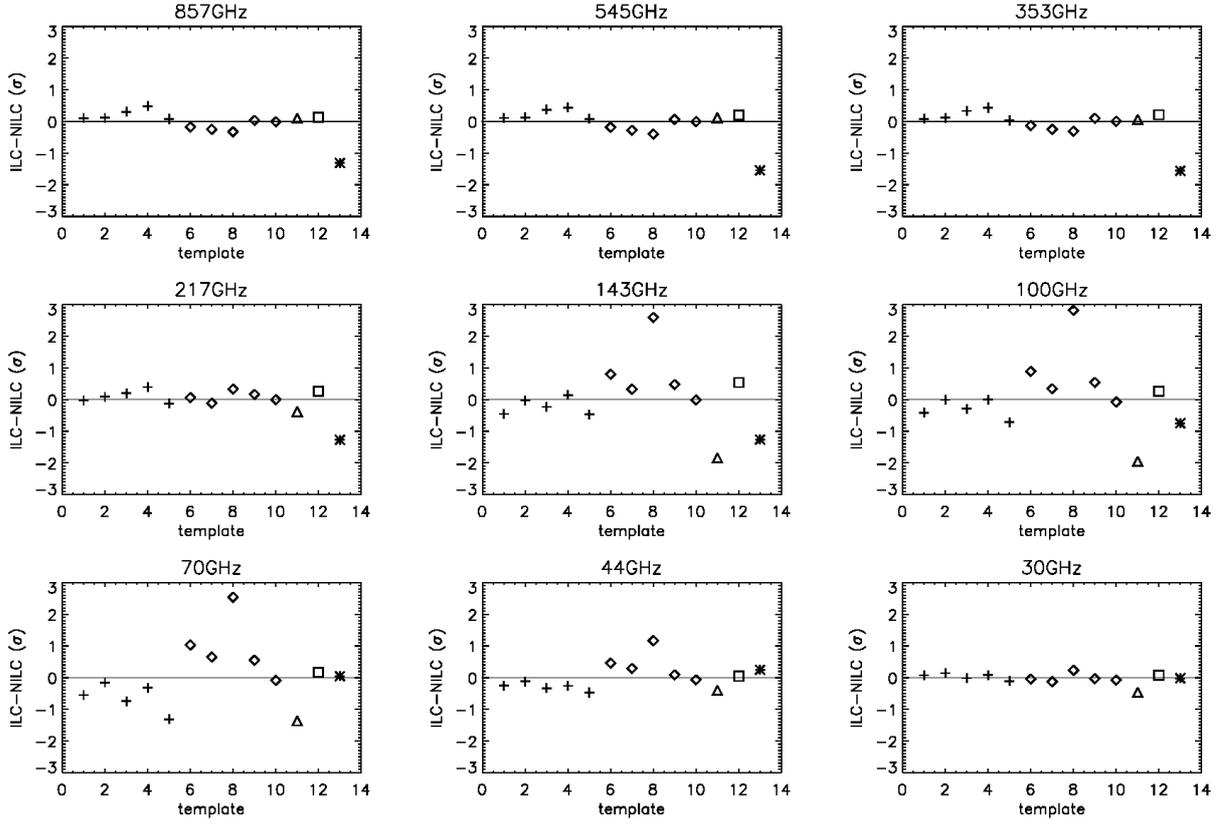}  
\caption{Comparison of emissivities derived after application of two different CMB subtractions. The value plotted is shown in Eq.~\ref{eq:sigdiff}. The 
symbols represent \ion{H}{i} (plus), H$_2$ (diamond), H$^{+}$ (triangle), Synchrotron (square) and dark gas (asterisk). Differences are generally small, but become larger 
in the range 143--44\,GHz.}
  \label{fig:emm_JA_cmb_per_Band}
\end{figure*}
\subsection{Effect of CMB subtraction}
\label{sec:CMB_rem}
The CMB removal in the \Planck\ maps we have used may create significant residuals in the Galactic plane. 
Therefore, to test the impact of the CMB removal on our results, a  different CMB estimate was reconstructed through a 
classical Internal Linear Combination (ILC) by means of Lagrange multipliers \citep{Eriksen2004}. A $\pm 15\degr$ 
strip in Galactic latitude was extracted from the HFI CMB frequency
channels maps (100, 143, 217 and 353\,GHz) reduced to a common resolution (10 arcmin), in units of K$_{\rm CMB}$. 
This CMB component was subtracted from all the \Planck\ LFI and HFI channels maps, resulting in a new and independent set of CMB-removed \Planck\ maps.

Even though the methods are similar, the resulting CMB in the Galactic plane is quite different. In constructing the 
 the standard CMB subtraction (Needlet ILC, ``NILC''), an intensity mask was applied \citep{planck2011-1.7} which results  
in CMB residuals remaining in a substantial part of the Galactic plane. No such mask was used in the Galactic plane ILC. 
Outside of this mask, the differences between the two CMB estimates are of the same order as the residuals of our fit at low frequencies, $44\GHz\, < \nu <143\GHz$.

Using these maps to perform the same inversion, we have concluded that the impact is small, and almost always within the error bars.
 In Fig.~\ref{fig:emm_JA_cmb}
we plot the derived emissivities using the NILC and the Galactic plane centred CMB removal (``ILC'').

To get a better idea of where the solution might be off by more than 1$\sigma$, we show the difference in units of $\sigma$ per band 
in Fig.~\ref{fig:emm_JA_cmb_per_Band}.
The value plotted is, for each component:
\begin{equation}
\label{eq:sigdiff}
  \Delta_{\sigma} = \frac{\epsilon_{\rm ILC} - \epsilon_{\rm NILC}}{\sqrt(\delta\epsilon_{\rm ILC}^2 + \delta\epsilon_{\rm NILC}^2) }\;.
\end{equation}

The agreement is generally very good, except for some points concentrated in the 
bands 143, 100 and 70 GHz. In these cases the CO templates are higher in the Galactic plane ILC
 whereas the \ion{H}{i} and H$^{+}$ templates are lower. Nevertheless, there are no points that are different by more than 3$\sigma$. Therefore, 
these differences do not change our main conclusions, but would slightly modify the amount of anomalous emission, for example.
As we are not modelling the specifics of any emission mechanisms at the lower 
frequencies, we are not sensitive to the impact of the CMB removal for this study.

\subsection{Impact of using dark gas template}

\begin{figure}
  \centering
 \includegraphics[width=0.45\linewidth]{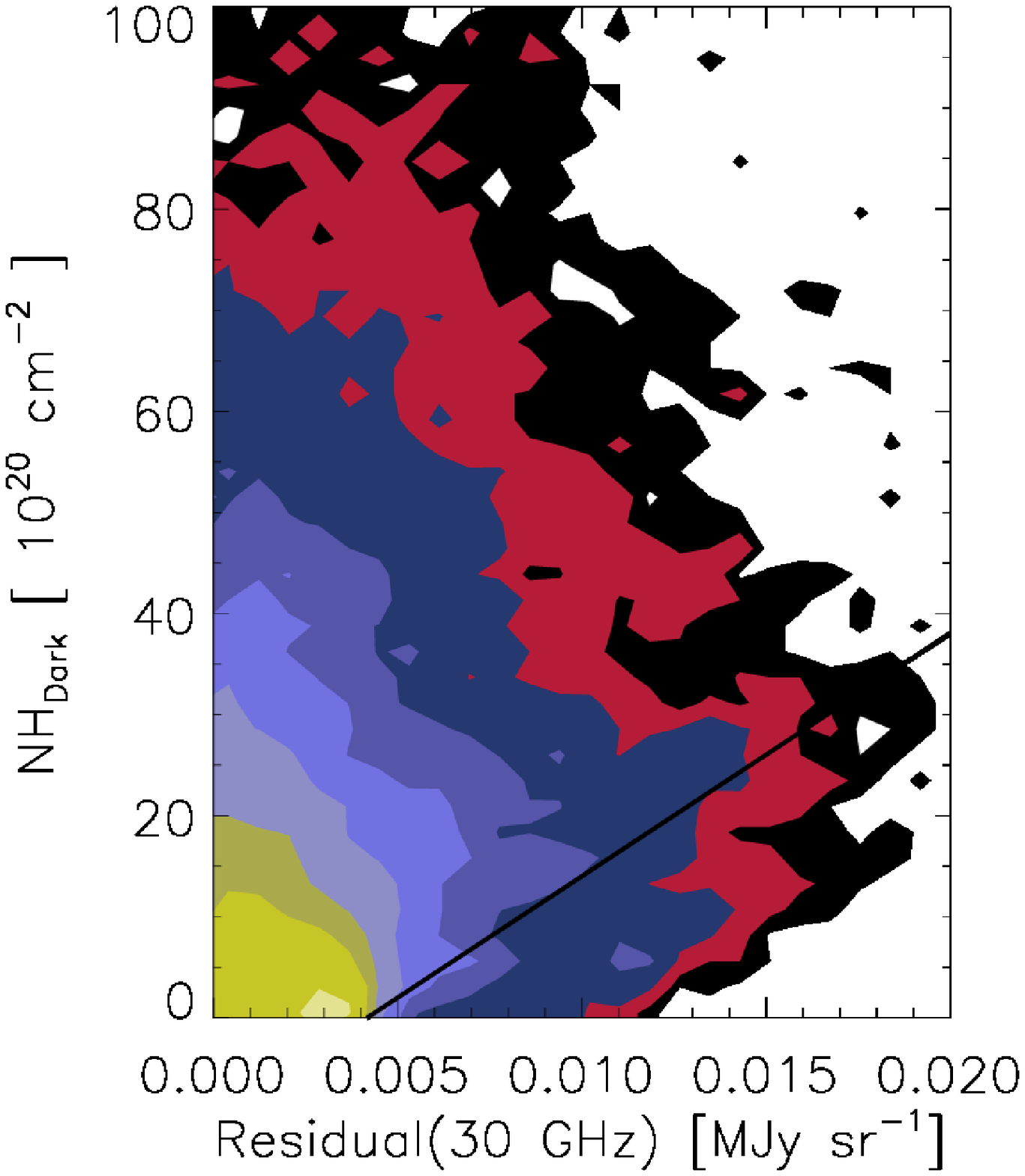}  
 \includegraphics[width=0.45\linewidth]{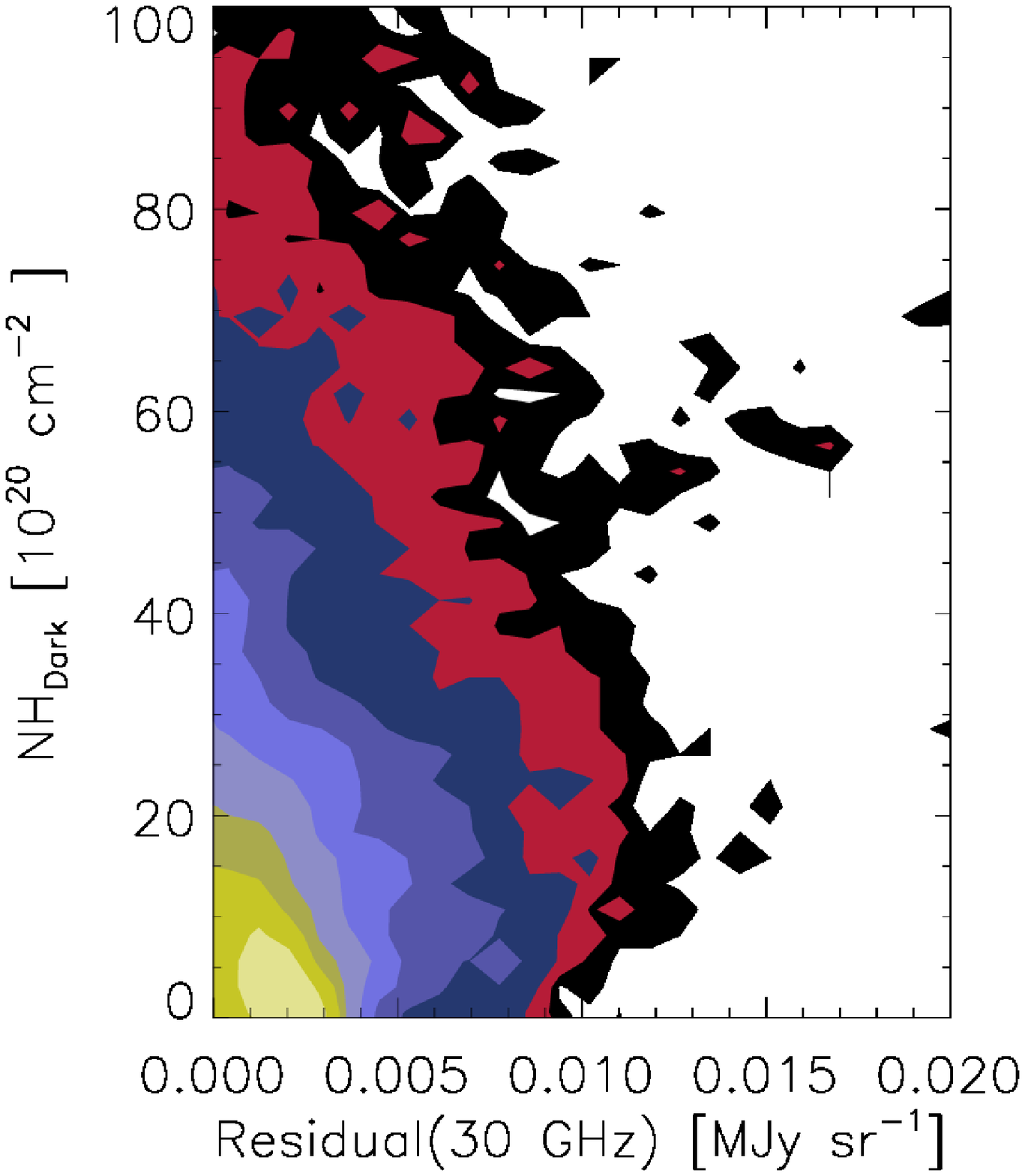}  
\caption{Dark gas intensity as a function of the residuals of the inversion at 30 GHz without (Left) and with (Right) the dark gas template. A distinct component
of the residual  (shown with the solid line) is seen to be correlated with the dark gas template.  
This correlated residual disappears when the dark gas is used in the inversion. }
  \label{fig:dg_vs_res}
\end{figure}

The dark gas template used in this study has been calculated by \cite{planck2011-7.0}. It is the result of finding at what column densities the correlation between 
total hydrogen column density (\ion{H}{i}+2$\ion{H}{_2}$) versus optical depth of the dust departs from linearity. The frequencies used to obtain the map at 857\,GHz that we 
use as a template are 857 and 545 GHz, along with the \textit{IRAS} $100\micron$ band. There is some circularity in using a template based on a residual in the \Planck\ data to 
fit the same data. However, the data used is restricted to the two highest HFI frequencies. Here we present the impact at 30 GHz, which in no respect suffers 
from the same circularity, as the data are far from the frequencies used to derive the dark gas template, and they come from LFI, a completely separate instrument from HFI.

We have performed the inversion with and without the dark gas template at 30 GHz, and looked at the correlation between the residuals and the 
column density of the dark gas. The results are shown in Fig.~\ref{fig:dg_vs_res}; the inversion performed without the dark gas template is on the left,
the same inversion with the dark gas template is on the right.
A distinct component of the residuals can be seen to correlate very well with the dark gas column density when the dark gas template is not used. This 
correlation completely disappears when it is used. In addition, the reduced $\chi^2$ for the fit shows a significant improvement of approximately $50\%$.

The choice of both the opacity correction for the \ion{H}{i} and the $X_{\rm CO}$ value chosen will have an impact on our results. We have chosen 
to keep the analysis simple for this study, as the spin temperature we assume, together with 
the dark gas template, should account for most of the optically thick gas in the plane. 
The ability of the dark gas template used to adequately account for all the 
optically thick gas remains to be studied, so future work could involve using more sophisticated methods to account for this effect.

\section{Conclusions}\label{sec:conclusion}

The analysis performed in this study 
provides a realistic description of the dust and gas properties as a function of Galactic radius in the Milky Way. 
The dust temperature in the \ion{H}{i} is seen to decrease as a function of Galactocentric distance from 24.0 to 13.9\,K, and the temperature in the molecular phase 
is heated by star formation associated to the Galactic spiral arms. 
The opacity found for grains in the solar circle is $\tau/N_{\rm H}=0.92\pm0.05\times10^{-25}\, {\rm cm}^2$, and with no significant variation 
with Galactic radius, even though the dust temperature is seen to drop by over 10\,K from the centre to the outer Galaxy.
The dust temperature in the \ion{H}{i} gas in the solar circle, $17.6\pm0.1$\,K is also compatible 
with the recent \Planck\ value of high-latitude cirrus ($17.8\pm0.9$\,K).

The extension of our analysis to lower frequencies than previously used in inversion techniques has allowed us to place constraints 
on the free-free, synchrotron and anomalous microwave emission:
\begin{itemize}
\item Anomalous dust emission is clearly 
seen in the atomic, molecular and dark gas phases. It is well fit by a very simple model consisting of spinning PAH molecules embedded in 
each of the gas phases.
 We highlight regions where spinning dust emission may be strong with respect to other emission mechanisms.
According to our simple model, in the Galactic plane spinning dust accounts for $25\pm5\%$(statistical) of the total emission at 30 GHz.
However, systematic uncertainties linked to our model may make this value more uncertain.
\item The dark gas phase has been explored spectrally across the Galactic plane for the first time. Its SED is similar to the molecular 
phase and therefore this phase seems to be tracing diffuse H$_2$ where CO is disassociated by energetic photons, as well as tracing 
optically thick CO and \ion{H}{i}.
\item The free-free emission is completely traced by our ionised component 
and indicates that the free-free estimate from {\it WMAP} component separation methods may be too high.
 In the ionised phase, 
spinning dust may be present but it is dominated by free-free emission.
\item Synchrotron emission is well characterised by a power law with a spectral index of $-1.0$. 
\end{itemize}

This first analysis of the large-scale   Galactic emission as seen by \Planck\ has constrained the large scale ISM properties in the Galaxy. 
It could be improved in the future via several modifications. Individual regions could be studied using higher angular resolution data, which would 
probe smaller scale fluctuations that have been missed by the present study.
The radial separation could be done differently to explore the difference between arm and inter-arm ISM, in order to focus more on the role of 
ISM conditions on the star formation process. It should also be possible to obtain an independent estimate of the dark gas in the plane, and 
ideally some radial information about its behaviour. In the near future, \Planck\ component separation products could be used in the place of 
{\it WMAP} ones. Use of radio recombination line surveys currently underway would be very useful to constrain the free-free emission 
in the plane, and as a function of Galactic radius.
\acknowledgements
A description of the Planck Collaboration and a list of its members can be found at \url{http://www.rssd.esa.int/index.php?project=PLANCK&page=Planck_Collaboration}

\bibliographystyle{aa}
\bibliography{16455}

\appendix

\section{Table of results}

\begin{table*}[b] 
\begingroup 
\newdimen\tblskip \tblskip=5pt
\caption{Result of the inversion per band and per component in MJy sr$^{-1}$ / 10$^{20}$ H atoms, except for the synchrotron component in which case it is in MJy sr$^{-1}$($\nu$) per MJy sr$^{-1}$(408 MHz).}
\label{tab:emm_res}
\nointerlineskip
\vskip -3mm
\footnotesize 
\setbox\tablebox=\vbox{ %
\newdimen\digitwidth 
\setbox0=\hbox{\rm 0}
\digitwidth=\wd0
\catcode`*=\active
\def*{\kern\digitwidth}
\newdimen\signwidth
\setbox0=\hbox{+}
\signwidth=\wd0
\catcode`!=\active
\newdimen\digitwidth 
\setbox0=\hbox{\rm 0}
\digitwidth=\wd0 
\catcode`*=\active 
\def*{\kern\digitwidth} 
\newdimen\signwidth 
\setbox0=\hbox{+} 
\signwidth=\wd0 
\catcode`!=\active 
\def!{\kern\signwidth} 

\newdimen\signwidth 
\setbox0=\hbox{.} 
\signwidth=\wd0 
\catcode`?=\active 
\def?{\kern\signwidth} 

\def\leaderfi1{\leaders\hbox to 5pt{\hss.\hss}\hfil}
{
\halign{\hbox to 0.7in{#\leaderfil}&
\hbox to 0.4in{\hfil#}&
#\hfil&
\hbox to 0.5in{#\hfil}&
\hbox to 0.4in{\hfil#}&
#\hfil&
\hbox to 0.5in{#\hfil}&
\hbox to 0.4in{\hfil#}&
#\hfil&
\hbox to 0.5in{#\hfil}&
\hbox to 0.4in{\hfil#}&
#\hfil&
\hbox to 0.5in{#\hfil}&
\hbox to 0.4in{\hfil#}&
#\hfil&
\hbox to 0.5in{#\hfil}\cr
\noalign{\vskip 2pt }
\noalign{\hrule \vskip 1pt }
\noalign{\hrule \vskip 1pt }
      Band [GHz] && $\ion{H}{i}(1)$ &&& $\ion{H}{i}(2)$ &&& $\ion{H}{i}(3)$ &&& $\ion{H}{i}(4)$ && $\ion{H}{i}(5)$ &\cr
\noalign{\hrule\vskip 3pt}
 25000& &1.4 $\pm$0.4&$\times 10^{-1}$ & & 5!* $\pm$3!*&$\times 10^{-2}$ & & 7.8 $\pm$0.6 &$\times 10^{-2}$ & & 2.4 $\pm$0.2 &$\times 10^{-2}$ & & 6.0 $\pm$0.9 &$\times 10^{-3}$\cr
 12000& $<$&1.5&$\times 10^{-1}$ & $<$& 4.6&$\times 10^{-2}$ & & 8.2 $\pm$1.3 &$\times 10^{-2}$ &  $<$&1.2 &$\times 10^{-2}$ &$<$& 5.9 &$\times 10^{-3}$\cr
  5000& &8?* $\pm$3?*&$\times 10^{-1}$ & & 3!* $\pm$2!*&$\times 10^{-1}$ & & 4.1 $\pm$0.5 &$\times 10^{-1}$ & $<$& 6.2 &$\times 10^{-2}$ &  $<$&2.5 &$\times 10^{-2}$\cr
  3000& &3.3 $\pm$1.5& & &2!* $\pm$1!*& &&1.8 $\pm$0.2 & &&5.9 $\pm$0.6 &$\times 10^{-1}$ &   $<$& 4.6 &$\times 10^{-2}$\cr
  2143& &4.9 $\pm$1.2& & &3.0 $\pm$0.8& &&2.6 $\pm$0.2 & &&1.3 $\pm$0.1 & &&2.6 $\pm$0.3 &$\times 10^{-1}$\cr
  1250& &1.5 $\pm$0.6& & &1.6 $\pm$0.4& && 1.5 $\pm$0.1 &&& 9.7 $\pm$0.4 &$\times 10^{-1}$ & & 4.0 $\pm$0.1 &$\times 10^{-1}$\cr
   857& &8!* $\pm$2!*&$\times 10^{-1}$ & & 9.2 $\pm$1.7&$\times 10^{-1}$ & & 7.7 $\pm$0.3 &$\times 10^{-1}$ & & 5.7 $\pm$0.1 &$\times 10^{-1}$ & & 3.2 $\pm$0.1 &$\times 10^{-1}$\cr
   545& &1.9 $\pm$0.7&$\times 10^{-1}$ & & 2.9 $\pm$0.5&$\times 10^{-1}$ & & 2.3 $\pm$0.1 &$\times 10^{-1}$ & & 2.0 $\pm$0.1 &$\times 10^{-1}$ & & 1.4 $\pm$0.1 &$\times 10^{-1}$\cr
   353& &3.8 $\pm$1.5&$\times 10^{-2}$ & & 6.7 $\pm$1.2&$\times 10^{-2}$ & & 5.4 $\pm$0.2 &$\times 10^{-2}$ & & 4.9 $\pm$0.1 &$\times 10^{-2}$ & & 4.0 $\pm$0.1 &$\times 10^{-2}$\cr
   217& &6!* $\pm$3!*&$\times 10^{-3}$ & & 1.4 $\pm$0.3&$\times 10^{-2}$ & & 1.1 $\pm$0.1 &$\times 10^{-2}$ & & 1.1 $\pm$0.1 &$\times 10^{-2}$ & & 9.5 $\pm$0.1 &$\times 10^{-3}$\cr
   143& &1.1 $\pm$0.7&$\times 10^{-3}$ & & 2.7 $\pm$0.5&$\times 10^{-3}$ & & 2.4 $\pm$0.1 &$\times 10^{-3}$ & & 2.3 $\pm$0.1 &$\times 10^{-3}$ & & 2.3 $\pm$0.1 &$\times 10^{-3}$\cr
   100& $<$&7.3&$\times 10^{-4}$ & & 6!* $\pm$2!*&$\times 10^{-4}$ & & 5.8 $\pm$0.6 &$\times 10^{-4}$ & & 5.9 $\pm$0.3 &$\times 10^{-4}$ & & 7.6 $\pm$0.1 &$\times 10^{-4}$\cr
    70& $<$&4.1&$\times 10^{-4}$ & & 1.1 $\pm$0.9&$\times 10^{-4}$ & & 1.6 $\pm$0.5 &$\times 10^{-4}$ & & 1.3 $\pm$0.1 &$\times 10^{-4}$ & & 2.9 $\pm$0.1 &$\times 10^{-4}$\cr
    44& $<$&3.3&$\times 10^{-4}$ & $<$& 8.9&$\times 10^{-5}$ & & 9!* $\pm$4!* &$\times 10^{-5}$ & & 8.3 $\pm$1.3 &$\times 10^{-5}$ & & 1.4 $\pm$0.2 &$\times 10^{-4}$\cr
    30& $<$&3.4&$\times 10^{-4}$ & & 1.5 $\pm$1.2&$\times 10^{-4}$ & & 1.3 $\pm$0.4 &$\times 10^{-4}$ & & 1.4 $\pm$0.1 &$\times 10^{-4}$ & & 1.2 $\pm$0.2 &$\times 10^{-4}$\cr
   1.4& $<$&1.1&$\times 10^{-3}$ &$<$& 3.8&$\times 10^{-4}$ &$<$& 3.1 &$\times 10^{-4}$ &$<$& 1.7 &$\times 10^{-4}$ & & 4!* $\pm$2!* &$\times 10^{-4}$\cr
\noalign{\vskip 5pt}
\noalign{\hrule \vskip 1pt}

      Band [GHz] && $\ion{H}{_2}(1)$ &&&$\ion{H}{_2}(2)$ &&&$\ion{H}{_2}(3)$ &&&$\ion{H}{_2}(4)$ &&$\ion{H}{_2}(5)$ &\cr
\noalign{\hrule\vskip 3pt}
 25000& &2.3 $\pm$1.3&$\times 10^{-2}$ & & 6.2 $\pm$0.6 &$\times 10^{-2}$ & & 9!* $\pm$6!* &$\times 10^{-3}$ & & 8!* $\pm$4!* &$\times 10^{-3}$ & & 3.0 $\pm$0.6 &$\times 10^{-2}$\cr
 12000& &4.8 $\pm$1.9&$\times 10^{-2}$ & & 5.1 $\pm$1.0 &$\times 10^{-2}$ & & 4.8 $\pm$1.1 &$\times 10^{-2}$ &$<$& 1.8 &$\times 10^{-2}$ &$<$& 3.6 &$\times 10^{-2}$\cr
  5000& &1.7 $\pm$1.3&$\times 10^{-1}$ & & 5.7 $\pm$0.7 &$\times 10^{-1}$ & & 2.5 $\pm$0.7 &$\times 10^{-1}$ &$<$& 8.1 &$\times 10^{-2}$ &$<$& 1.8 &$\times 10^{-1}$\cr
  3000&&1.4 $\pm$0.6 & &&2.3 $\pm$0.3 & & & 8!* $\pm$2!* &$\times 10^{-1}$ & & 2.7 $\pm$1.0 &$\times 10^{-1}$ & & 9!* $\pm$2!* &$\times 10^{-1}$\cr
  2143&&2.8 $\pm$0.4 & &&3.6 $\pm$0.2 & && 1.8 $\pm$0.2 & & & 4.3 $\pm$0.9 &$\times 10^{-1}$ & &1.5 $\pm$0.3 &$\times 10^{  0}$\cr
  1250&&2.0 $\pm$0.2 & &&2.1 $\pm$0.1 & && 1.3 $\pm$0.1 & & & 5.3 $\pm$0.4 &$\times 10^{-1}$ & & 1.1 $\pm$0.1 &$\times 10^{  0}$\cr
   857& &8.0 $\pm$0.8&$\times 10^{-1}$ & & 8.3 $\pm$0.4 &$\times 10^{-1}$ & & 6.1 $\pm$0.4 &$\times 10^{-1}$ & & 3.8 $\pm$0.2 &$\times 10^{-1}$ & & 6.1 $\pm$0.5 &$\times 10^{-1}$\cr
   545& &2.1 $\pm$0.2&$\times 10^{-1}$ & & 2.1 $\pm$0.1 &$\times 10^{-1}$ & & 1.7 $\pm$0.1 &$\times 10^{-1}$ & & 1.4 $\pm$0.1 &$\times 10^{-1}$ & & 2.1 $\pm$0.1 &$\times 10^{-1}$\cr
   353& &4.3 $\pm$0.5&$\times 10^{-2}$ & & 4.1 $\pm$0.3 &$\times 10^{-2}$ & & 3.7 $\pm$0.2 &$\times 10^{-2}$ & & 3.7 $\pm$0.1 &$\times 10^{-2}$ & & 5.3 $\pm$0.3 &$\times 10^{-2}$\cr
   217& &1.1 $\pm$0.1&$\times 10^{-2}$ & & 1.0 $\pm$0.1 &$\times 10^{-2}$ & & 9.5 $\pm$0.4 &$\times 10^{-3}$ & & 1.0 $\pm$0.1 &$\times 10^{-2}$ & & 1.4 $\pm$0.1 &$\times 10^{-2}$\cr
   143& &1.4 $\pm$0.2&$\times 10^{-3}$ & & 1.3 $\pm$0.1 &$\times 10^{-3}$ & & 1.2 $\pm$0.1 &$\times 10^{-3}$ & & 1.7 $\pm$0.1 &$\times 10^{-3}$ & & 2.3 $\pm$0.2 &$\times 10^{-3}$\cr
   100& &1.9 $\pm$0.1&$\times 10^{-3}$ & & 1.6 $\pm$0.1 &$\times 10^{-3}$ & & 1.5 $\pm$0.1 &$\times 10^{-3}$ & & 1.6 $\pm$0.1 &$\times 10^{-3}$ & & 1.7 $\pm$0.2 &$\times 10^{-3}$\cr
    70& &2.1 $\pm$0.5&$\times 10^{-4}$ & & 1.8 $\pm$0.2 &$\times 10^{-4}$ & & 2.2 $\pm$0.3 &$\times 10^{-4}$ & & 2.3 $\pm$0.1 &$\times 10^{-4}$ & & 3.7 $\pm$1.6 &$\times 10^{-4}$\cr
    44& &1.9 $\pm$0.4&$\times 10^{-4}$ & & 1.7 $\pm$0.3 &$\times 10^{-4}$ & & 2.4 $\pm$0.3 &$\times 10^{-4}$ & & 1.3 $\pm$0.1 &$\times 10^{-4}$ & & 4.4 $\pm$1.9 &$\times 10^{-4}$\cr
    30& &2.8 $\pm$0.4&$\times 10^{-4}$ & & 2.5 $\pm$0.4 &$\times 10^{-4}$ & & 3.1 $\pm$0.3 &$\times 10^{-4}$ & & 1.3 $\pm$0.1 &$\times 10^{-4}$ & & 6!* $\pm$2!* &$\times 10^{-4}$\cr
   1.4&$<$&3.6&$\times 10^{-4}$ &$<$& 1.6 &$\times 10^{-4}$ &$<$& 1.7 &$\times 10^{-4}$ &$<$& 8.5 &$\times 10^{-5}$ &$<$& 5.9 &$\times 10^{-3}$\cr
\noalign{\vskip 5pt}
\noalign{\hrule \vskip 1pt}
      Band [GHz]& & $\ion{H}{ii}$ &&&Sync. &&& Dark &&&&\cr
\noalign{\hrule\vskip 3pt}
 25000& &9.7 $\pm$0.9&$\times 10^{-1 }$&& $-$ & & & 4!* $\pm$3!* &$\times 10^{-3}$& & \cr
 12000& &3.3 $\pm$0.3&& & $-$& &$<$& 1.5 &$\times 10^{-2}$& & \cr
  5000& &1.9 $\pm$0.1&$\times 10^{ 1 }$& &$-$& &$<$& 7.1 &$\times 10^{-2}$& & \cr
  3000& &2.8 $\pm$0.1&$\times 10^{ 1 }$& & $-$& &$<$& 1.7 &$\times 10^{-1}$& & \cr
  2143& &2.9 $\pm$0.1&$\times 10^{ 1 }$& &$-$& & & 1.0 $\pm$0.1 && & \cr
  1250& &1.2 $\pm$0.1&$\times 10^{ 1 }$& & $-$& & & 8.3 $\pm$0.4 &$\times 10^{-1}$& & \cr
   857& &4.7 $\pm$0.2&& &$-$& & & 4.4 $\pm$0.2 &$\times 10^{-1}$& & \cr
   545& &1.3 $\pm$0.1&& &$-$& & & 1.5 $\pm$0.1 &$\times 10^{-1}$& & \cr
   353& &3.1 $\pm$0.1&$\times 10^{-1 }$&&$-$& & & 3.5 $\pm$0.1 &$\times 10^{-2}$& & \cr
   217& &7.9 $\pm$0.2&$\times 10^{-2 }$&&$-$& & & 7.9 $\pm$0.2 &$\times 10^{-3}$& & \cr
   143& &3.0 $\pm$0.1&$\times 10^{-2 }$&& $-$ & & & 1.8 $\pm$0.1 &$\times 10^{-3}$& & \cr
   100& &2.3 $\pm$0.1&$\times 10^{-2 }$& &3!* $\pm$2!* &$\times 10^{-3 }$& &7.1 $\pm$0.3 &$\times 10^{-4}$& & \cr
    70& &2.2 $\pm$0.1&$\times 10^{-2 }$&$<$&4.7 &$\times 10^{-3 }$&& 2.5 $\pm$0.2 &$\times 10^{-4}$& & \cr
    44& &2.3 $\pm$0.1&$\times 10^{-2 }$& &9!* $\pm$3!* &$\times 10^{-3 }$& &1.5 $\pm$0.2 &$\times 10^{-4}$& & \cr
    30& &2.2 $\pm$0.1&$\times 10^{-2 }$& &2.0 $\pm$0.3 &$\times 10^{-2 }$& &1.7 $\pm$0.2 &$\times 10^{-4}$& & \cr
     1.4& &1.8 $\pm$0.1&$\times 10^{-2 }$& &2.9 $\pm$0.1 &$\times 10^{-1 }$&$<$& 3.5  &$\times 10^{-5}$& & \cr
\noalign{\vskip 2pt}
\noalign{\vskip 5pt\hrule\vskip 3pt}}}}
\endPlancktable 
\endgroup
\end{table*}
\twocolumn
\end{document}

%% file: 16455_a.tex
\def\setsymbol#1#2{\expandafter\def\csname #1\endcsname{#2}}
\def\getsymbol#1{\csname #1\endcsname}

\def\Planck{{\it Planck\/}}

\newbox\tablebox    \newdimen\tablewidth
\def\leaderfil{\leaders\hbox to 5pt{\hss.\hss}\hfil}
\def\endPlancktable{\tablewidth=\columnwidth 
    $$\hss\copy\tablebox\hss$$
    \vskip-\lastskip\vskip -2pt}

\def\tablenote#1 #2\par{\begingroup \parindent=0.8em
    \abovedisplayshortskip=0pt\belowdisplayshortskip=0pt
    \noindent
    $$\hss\vbox{\hsize\tablewidth \hangindent=\parindent \hangafter=1 \noindent
    \hbox to \parindent{\sup{\rm #1}\hss}\strut#2\strut\par}\hss$$
    \endgroup}

\def\L2{\ifmmode L_2\else $L_2$\fi}

\def\DeltaT{\ifmmode \Delta T\else $\Delta T$\fi}
\def\deltat{\ifmmode \Delta t\else $\Delta t$\fi}
\def\fknee{\ifmmode f_{\rm knee}\else $f_{\rm knee}$\fi}
\def\Fmax{\ifmmode F_{\rm max}\else $F_{\rm max}$\fi}
\def\solar{\ifmmode{\rm M}_{\mathord\odot}\else${\rm M}_{\mathord\odot}$\fi}

\def\inv{\ifmmode^{-1}\else$^{-1}$\fi}
\def\mo{\ifmmode^{-1}\else$^{-1}$\fi}
\def\sup#1{\ifmmode ^{\rm #1}\else $^{\rm #1}$\fi}
\def\expo#1{\ifmmode \times 10^{#1}\else $\times 10^{#1}$\fi}
\def\,{\thinspace}
\def\lsim{\mathrel{\raise .4ex\hbox{\rlap{$<$}\lower 1.2ex\hbox{$\sim$}}}}
\def\gsim{\mathrel{\raise .4ex\hbox{\rlap{$>$}\lower 1.2ex\hbox{$\sim$}}}}

\def\simprop{\mathrel{\raise .4ex\hbox{\rlap{$\propto$}\lower 1.2ex\hbox{$\sim$}}}}
\def\deg{\ifmmode^\circ\else$^\circ$\fi}
\def\pdeg{\ifmmode $\setbox0=\hbox{$^{\circ}$}\rlap{\hskip.11\wd0 .}$^{\circ}
          \else \setbox0=\hbox{$^{\circ}$}\rlap{\hskip.11\wd0 .}$^{\circ}$\fi}
\def\arcs{\ifmmode {^{\scriptstyle\prime\prime}}
          \else $^{\scriptstyle\prime\prime}$\fi}
\def\arcm{\ifmmode {^{\scriptstyle\prime}}
          \else $^{\scriptstyle\prime}$\fi}
\newdimen\sa  \newdimen\sb
\def\parcs{\sa=.07em \sb=.03em
     \ifmmode \hbox{\rlap{.}}^{\scriptstyle\prime\kern -\sb\prime}\hbox{\kern -\sa}
     \else \rlap{.}$^{\scriptstyle\prime\kern -\sb\prime}$\kern -\sa\fi}
\def\parcm{\sa=.08em \sb=.03em
     \ifmmode \hbox{\rlap{.}\kern\sa}^{\scriptstyle\prime}\hbox{\kern-\sb}
     \else \rlap{.}\kern\sa$^{\scriptstyle\prime}$\kern-\sb\fi}
\def\ra[#1 #2 #3.#4]{#1\sup{h}#2\sup{m}#3\sup{s}\llap.#4}
\def\dec[#1 #2 #3.#4]{#1\deg#2\arcm#3\arcs\llap.#4}
\def\deco[#1 #2 #3]{#1\deg#2\arcm#3\arcs}
\def\rra[#1 #2]{#1\sup{h}#2\sup{m}}

\def\dots{\relax\ifmmode \ldots\else $\ldots$\fi}
\def\WHzsr{\ifmmode $W\,Hz\mo\,sr\mo$\else W\,Hz\mo\,sr\mo\fi}
\def\mHz{\ifmmode $\,mHz$\else \,mHz\fi}
\def\GHz{\ifmmode $\,GHz$\else \,GHz\fi}
\def\mKs{\ifmmode $\,mK\,s$^{1/2}\else \,mK\,s$^{1/2}$\fi}
\def\muKs{\ifmmode \,\mu$K\,s$^{1/2}\else \,$\mu$K\,s$^{1/2}$\fi}
\def\muKRJs{\ifmmode \,\mu$K$_{\rm RJ}$\,s$^{1/2}\else \,$\mu$K$_{\rm RJ}$\,s$^{1/2}$\fi}
\def\muKHz{\ifmmode \,\mu$K\,Hz$^{-1/2}\else \,$\mu$K\,Hz$^{-1/2}$\fi}
\def\MJysr{\ifmmode \,$MJy\,sr\mo$\else \,MJy\,sr\mo\fi}
\def\MJysrmK{\ifmmode \,$MJy\,sr\mo$\,mK$_{\rm CMB}\mo\else \,MJy\,sr\mo\,mK$_{\rm CMB}\mo$\fi}
\def\microns{\ifmmode \,\mu$m$\else \,$\mu$m\fi}
\def\micron{\microns}
\def\muK{\ifmmode \,\mu$K$\else \,$\mu$\hbox{K}\fi}
\def\microK{\ifmmode \,\mu$K$\else \,$\mu$\hbox{K}\fi}
\def\muW{\ifmmode \,\mu$W$\else \,$\mu$\hbox{W}\fi}
\def\kms{\ifmmode $\,km\,s$^{-1}\else \,km\,s$^{-1}$\fi}
\def\kmsMpc{\ifmmode $\,\kms\,Mpc\mo$\else \,\kms\,Mpc\mo\fi}
\setsymbol{LFI:center:frequency:70GHz:units}{70.3\,GHz}
\setsymbol{LFI:center:frequency:44GHz:units}{44.1\,GHz}
\setsymbol{LFI:center:frequency:30GHz:units}{28.5\,GHz}

\setsymbol{LFI:center:frequency:70GHz}{70.3}
\setsymbol{LFI:center:frequency:44GHz}{44.1}
\setsymbol{LFI:center:frequency:30GHz}{28.5}

\setsymbol{LFI:center:frequency:LFI18:Rad:M:units}{71.7\GHz}
\setsymbol{LFI:center:frequency:LFI19:Rad:M:units}{67.5\GHz}
\setsymbol{LFI:center:frequency:LFI20:Rad:M:units}{69.2\GHz}
\setsymbol{LFI:center:frequency:LFI21:Rad:M:units}{70.4\GHz}
\setsymbol{LFI:center:frequency:LFI22:Rad:M:units}{71.5\GHz}
\setsymbol{LFI:center:frequency:LFI23:Rad:M:units}{70.8\GHz}
\setsymbol{LFI:center:frequency:LFI24:Rad:M:units}{44.4\GHz}
\setsymbol{LFI:center:frequency:LFI25:Rad:M:units}{44.0\GHz}
\setsymbol{LFI:center:frequency:LFI26:Rad:M:units}{43.9\GHz}
\setsymbol{LFI:center:frequency:LFI27:Rad:M:units}{28.3\GHz}
\setsymbol{LFI:center:frequency:LFI28:Rad:M:units}{28.8\GHz}
\setsymbol{LFI:center:frequency:LFI18:Rad:S:units}{70.1\GHz}
\setsymbol{LFI:center:frequency:LFI19:Rad:S:units}{69.6\GHz}
\setsymbol{LFI:center:frequency:LFI20:Rad:S:units}{69.5\GHz}
\setsymbol{LFI:center:frequency:LFI21:Rad:S:units}{69.5\GHz}
\setsymbol{LFI:center:frequency:LFI22:Rad:S:units}{72.8\GHz}
\setsymbol{LFI:center:frequency:LFI23:Rad:S:units}{71.3\GHz}
\setsymbol{LFI:center:frequency:LFI24:Rad:S:units}{44.1\GHz}
\setsymbol{LFI:center:frequency:LFI25:Rad:S:units}{44.1\GHz}
\setsymbol{LFI:center:frequency:LFI26:Rad:S:units}{44.1\GHz}
\setsymbol{LFI:center:frequency:LFI27:Rad:S:units}{28.5\GHz}
\setsymbol{LFI:center:frequency:LFI28:Rad:S:units}{28.2\GHz}

\setsymbol{LFI:center:frequency:LFI18:Rad:M}{71.7}
\setsymbol{LFI:center:frequency:LFI19:Rad:M}{67.5}
\setsymbol{LFI:center:frequency:LFI20:Rad:M}{69.2}
\setsymbol{LFI:center:frequency:LFI21:Rad:M}{70.4}
\setsymbol{LFI:center:frequency:LFI22:Rad:M}{71.5}
\setsymbol{LFI:center:frequency:LFI23:Rad:M}{70.8}
\setsymbol{LFI:center:frequency:LFI24:Rad:M}{44.4}
\setsymbol{LFI:center:frequency:LFI25:Rad:M}{44.0}
\setsymbol{LFI:center:frequency:LFI26:Rad:M}{43.9}
\setsymbol{LFI:center:frequency:LFI27:Rad:M}{28.3}
\setsymbol{LFI:center:frequency:LFI28:Rad:M}{28.8}
\setsymbol{LFI:center:frequency:LFI18:Rad:S}{70.1}
\setsymbol{LFI:center:frequency:LFI19:Rad:S}{69.6}
\setsymbol{LFI:center:frequency:LFI20:Rad:S}{69.5}
\setsymbol{LFI:center:frequency:LFI21:Rad:S}{69.5}
\setsymbol{LFI:center:frequency:LFI22:Rad:S}{72.8}
\setsymbol{LFI:center:frequency:LFI23:Rad:S}{71.3}
\setsymbol{LFI:center:frequency:LFI24:Rad:S}{44.1}
\setsymbol{LFI:center:frequency:LFI25:Rad:S}{44.1}
\setsymbol{LFI:center:frequency:LFI26:Rad:S}{44.1}
\setsymbol{LFI:center:frequency:LFI27:Rad:S}{28.5}
\setsymbol{LFI:center:frequency:LFI28:Rad:S}{28.2}

\setsymbol{LFI:white:noise:sensitivity:70GHz:units}{134.7\muKs}
\setsymbol{LFI:white:noise:sensitivity:44GHz:units}{164.7\muKs}
\setsymbol{LFI:white:noise:sensitivity:30GHz:units}{143.4\muKs}

\setsymbol{LFI:white:noise:sensitivity:70GHz}{134.7}
\setsymbol{LFI:white:noise:sensitivity:44GHz}{164.7}
\setsymbol{LFI:white:noise:sensitivity:30GHz}{143.4}

\setsymbol{LFI:white:noise:sensitivity:LFI18:Rad:M:units}{512.0\muKs}
\setsymbol{LFI:white:noise:sensitivity:LFI19:Rad:M:units}{581.4\muKs}
\setsymbol{LFI:white:noise:sensitivity:LFI20:Rad:M:units}{590.8\muKs}
\setsymbol{LFI:white:noise:sensitivity:LFI21:Rad:M:units}{455.2\muKs}
\setsymbol{LFI:white:noise:sensitivity:LFI22:Rad:M:units}{492.0\muKs}
\setsymbol{LFI:white:noise:sensitivity:LFI23:Rad:M:units}{507.7\muKs}
\setsymbol{LFI:white:noise:sensitivity:LFI24:Rad:M:units}{462.2\muKs}
\setsymbol{LFI:white:noise:sensitivity:LFI25:Rad:M:units}{413.6\muKs}
\setsymbol{LFI:white:noise:sensitivity:LFI26:Rad:M:units}{478.6\muKs}
\setsymbol{LFI:white:noise:sensitivity:LFI27:Rad:M:units}{277.7\muKs}
\setsymbol{LFI:white:noise:sensitivity:LFI28:Rad:M:units}{312.3\muKs}
\setsymbol{LFI:white:noise:sensitivity:LFI18:Rad:S:units}{465.7\muKs}
\setsymbol{LFI:white:noise:sensitivity:LFI19:Rad:S:units}{555.6\muKs}
\setsymbol{LFI:white:noise:sensitivity:LFI20:Rad:S:units}{623.2\muKs}
\setsymbol{LFI:white:noise:sensitivity:LFI21:Rad:S:units}{564.1\muKs}
\setsymbol{LFI:white:noise:sensitivity:LFI22:Rad:S:units}{534.4\muKs}
\setsymbol{LFI:white:noise:sensitivity:LFI23:Rad:S:units}{542.4\muKs}
\setsymbol{LFI:white:noise:sensitivity:LFI24:Rad:S:units}{399.2\muKs}
\setsymbol{LFI:white:noise:sensitivity:LFI25:Rad:S:units}{392.6\muKs}
\setsymbol{LFI:white:noise:sensitivity:LFI26:Rad:S:units}{418.6\muKs}
\setsymbol{LFI:white:noise:sensitivity:LFI27:Rad:S:units}{302.9\muKs}
\setsymbol{LFI:white:noise:sensitivity:LFI28:Rad:S:units}{285.3\muKs}

\setsymbol{LFI:white:noise:sensitivity:LFI18:Rad:M}{512.0}
\setsymbol{LFI:white:noise:sensitivity:LFI19:Rad:M}{581.4}
\setsymbol{LFI:white:noise:sensitivity:LFI20:Rad:M}{590.8}
\setsymbol{LFI:white:noise:sensitivity:LFI21:Rad:M}{455.2}
\setsymbol{LFI:white:noise:sensitivity:LFI22:Rad:M}{492.0}
\setsymbol{LFI:white:noise:sensitivity:LFI23:Rad:M}{507.7}
\setsymbol{LFI:white:noise:sensitivity:LFI24:Rad:M}{462.2}
\setsymbol{LFI:white:noise:sensitivity:LFI25:Rad:M}{413.6}
\setsymbol{LFI:white:noise:sensitivity:LFI26:Rad:M}{478.6}
\setsymbol{LFI:white:noise:sensitivity:LFI27:Rad:M}{277.7}
\setsymbol{LFI:white:noise:sensitivity:LFI28:Rad:M}{312.3}
\setsymbol{LFI:white:noise:sensitivity:LFI18:Rad:S}{465.7}
\setsymbol{LFI:white:noise:sensitivity:LFI19:Rad:S}{555.6}
\setsymbol{LFI:white:noise:sensitivity:LFI20:Rad:S}{623.2}
\setsymbol{LFI:white:noise:sensitivity:LFI21:Rad:S}{564.1}
\setsymbol{LFI:white:noise:sensitivity:LFI22:Rad:S}{534.4}
\setsymbol{LFI:white:noise:sensitivity:LFI23:Rad:S}{542.4}
\setsymbol{LFI:white:noise:sensitivity:LFI24:Rad:S}{399.2}
\setsymbol{LFI:white:noise:sensitivity:LFI25:Rad:S}{392.6}
\setsymbol{LFI:white:noise:sensitivity:LFI26:Rad:S}{418.6}
\setsymbol{LFI:white:noise:sensitivity:LFI27:Rad:S}{302.9}
\setsymbol{LFI:white:noise:sensitivity:LFI28:Rad:S}{285.3}

\setsymbol{LFI:knee:frequency:70GHz:units}{29.5\mHz}
\setsymbol{LFI:knee:frequency:44GHz:units}{56.2\mHz}
\setsymbol{LFI:knee:frequency:30GHz:units}{113.7\mHz}

\setsymbol{LFI:knee:frequency:70GHz}{29.5}
\setsymbol{LFI:knee:frequency:44GHz}{56.2}
\setsymbol{LFI:knee:frequency:30GHz}{113.7}

\setsymbol{LFI:knee:frequency:LFI18:Rad:M:units}{16.3\mHz}
\setsymbol{LFI:knee:frequency:LFI19:Rad:M:units}{15.1\mHz}
\setsymbol{LFI:knee:frequency:LFI20:Rad:M:units}{18.7\mHz}
\setsymbol{LFI:knee:frequency:LFI21:Rad:M:units}{37.2\mHz}
\setsymbol{LFI:knee:frequency:LFI22:Rad:M:units}{12.7\mHz}
\setsymbol{LFI:knee:frequency:LFI23:Rad:M:units}{34.6\mHz}
\setsymbol{LFI:knee:frequency:LFI24:Rad:M:units}{46.2\mHz}
\setsymbol{LFI:knee:frequency:LFI25:Rad:M:units}{24.9\mHz}
\setsymbol{LFI:knee:frequency:LFI26:Rad:M:units}{67.6\mHz}
\setsymbol{LFI:knee:frequency:LFI27:Rad:M:units}{187.4\mHz}
\setsymbol{LFI:knee:frequency:LFI28:Rad:M:units}{122.2\mHz}
\setsymbol{LFI:knee:frequency:LFI18:Rad:S:units}{17.7\mHz}
\setsymbol{LFI:knee:frequency:LFI19:Rad:S:units}{22.0\mHz}
\setsymbol{LFI:knee:frequency:LFI20:Rad:S:units}{8.7\mHz}
\setsymbol{LFI:knee:frequency:LFI21:Rad:S:units}{25.9\mHz}
\setsymbol{LFI:knee:frequency:LFI22:Rad:S:units}{15.8\mHz}
\setsymbol{LFI:knee:frequency:LFI23:Rad:S:units}{129.8\mHz}
\setsymbol{LFI:knee:frequency:LFI24:Rad:S:units}{100.9\mHz}
\setsymbol{LFI:knee:frequency:LFI25:Rad:S:units}{38.9\mHz}
\setsymbol{LFI:knee:frequency:LFI26:Rad:S:units}{58.9\mHz}
\setsymbol{LFI:knee:frequency:LFI27:Rad:S:units}{104.4\mHz}
\setsymbol{LFI:knee:frequency:LFI28:Rad:S:units}{40.7\mHz}

\setsymbol{LFI:knee:frequency:LFI18:Rad:M}{16.3}
\setsymbol{LFI:knee:frequency:LFI19:Rad:M}{15.1}
\setsymbol{LFI:knee:frequency:LFI20:Rad:M}{18.7}
\setsymbol{LFI:knee:frequency:LFI21:Rad:M}{37.2}
\setsymbol{LFI:knee:frequency:LFI22:Rad:M}{12.7}
\setsymbol{LFI:knee:frequency:LFI23:Rad:M}{34.6}
\setsymbol{LFI:knee:frequency:LFI24:Rad:M}{46.2}
\setsymbol{LFI:knee:frequency:LFI25:Rad:M}{24.9}
\setsymbol{LFI:knee:frequency:LFI26:Rad:M}{67.6}
\setsymbol{LFI:knee:frequency:LFI27:Rad:M}{187.4}
\setsymbol{LFI:knee:frequency:LFI28:Rad:M}{122.2}
\setsymbol{LFI:knee:frequency:LFI18:Rad:S}{17.7}
\setsymbol{LFI:knee:frequency:LFI19:Rad:S}{22.0}
\setsymbol{LFI:knee:frequency:LFI20:Rad:S}{8.7}
\setsymbol{LFI:knee:frequency:LFI21:Rad:S}{25.9}
\setsymbol{LFI:knee:frequency:LFI22:Rad:S}{15.8}
\setsymbol{LFI:knee:frequency:LFI23:Rad:S}{129.8}
\setsymbol{LFI:knee:frequency:LFI24:Rad:S}{100.9}
\setsymbol{LFI:knee:frequency:LFI25:Rad:S}{38.9}
\setsymbol{LFI:knee:frequency:LFI26:Rad:S}{58.9}
\setsymbol{LFI:knee:frequency:LFI27:Rad:S}{104.4}
\setsymbol{LFI:knee:frequency:LFI28:Rad:S}{40.7}

\setsymbol{LFI:slope:70GHz:units}{$-1.03$\mHz}
\setsymbol{LFI:slope:44GHz:units}{$-0.89$\mHz}
\setsymbol{LFI:slope:30GHz:units}{$-0.87$\mHz}

\setsymbol{LFI:slope:70GHz}{$-1.03$}
\setsymbol{LFI:slope:44GHz}{$-0.89$}
\setsymbol{LFI:slope:30GHz}{$-0.87$}

\setsymbol{LFI:slope:LFI18:Rad:M:units}{$-1.04$\mHz}
\setsymbol{LFI:slope:LFI19:Rad:M:units}{$-1.09$\mHz}
\setsymbol{LFI:slope:LFI20:Rad:M:units}{$-0.69$\mHz}
\setsymbol{LFI:slope:LFI21:Rad:M:units}{$-1.56$\mHz}
\setsymbol{LFI:slope:LFI22:Rad:M:units}{$-1.01$\mHz}
\setsymbol{LFI:slope:LFI23:Rad:M:units}{$-0.96$\mHz}
\setsymbol{LFI:slope:LFI24:Rad:M:units}{$-0.83$\mHz}
\setsymbol{LFI:slope:LFI25:Rad:M:units}{$-0.91$\mHz}
\setsymbol{LFI:slope:LFI26:Rad:M:units}{$-0.95$\mHz}
\setsymbol{LFI:slope:LFI27:Rad:M:units}{$-0.87$\mHz}
\setsymbol{LFI:slope:LFI28:Rad:M:units}{$-0.88$\mHz}
\setsymbol{LFI:slope:LFI18:Rad:S:units}{$-1.15$\mHz}
\setsymbol{LFI:slope:LFI19:Rad:S:units}{$-1.00$\mHz}
\setsymbol{LFI:slope:LFI20:Rad:S:units}{$-0.95$\mHz}
\setsymbol{LFI:slope:LFI21:Rad:S:units}{$-0.92$\mHz}
\setsymbol{LFI:slope:LFI22:Rad:S:units}{$-1.01$\mHz}
\setsymbol{LFI:slope:LFI23:Rad:S:units}{$-0.95$\mHz}
\setsymbol{LFI:slope:LFI24:Rad:S:units}{$-0.73$\mHz}
\setsymbol{LFI:slope:LFI25:Rad:S:units}{$-1.16$\mHz}
\setsymbol{LFI:slope:LFI26:Rad:S:units}{$-0.79$\mHz}
\setsymbol{LFI:slope:LFI27:Rad:S:units}{$-0.82$\mHz}
\setsymbol{LFI:slope:LFI28:Rad:S:units}{$-0.91$\mHz}

\setsymbol{LFI:slope:LFI18:Rad:M}{$-1.04$}
\setsymbol{LFI:slope:LFI19:Rad:M}{$-1.09$}
\setsymbol{LFI:slope:LFI20:Rad:M}{$-0.69$}
\setsymbol{LFI:slope:LFI21:Rad:M}{$-1.56$}
\setsymbol{LFI:slope:LFI22:Rad:M}{$-1.01$}
\setsymbol{LFI:slope:LFI23:Rad:M}{$-0.96$}
\setsymbol{LFI:slope:LFI24:Rad:M}{$-0.83$}
\setsymbol{LFI:slope:LFI25:Rad:M}{$-0.91$}
\setsymbol{LFI:slope:LFI26:Rad:M}{$-0.95$}
\setsymbol{LFI:slope:LFI27:Rad:M}{$-0.87$}
\setsymbol{LFI:slope:LFI28:Rad:M}{$-0.88$}
\setsymbol{LFI:slope:LFI18:Rad:S}{$-1.15$}
\setsymbol{LFI:slope:LFI19:Rad:S}{$-1.00$}
\setsymbol{LFI:slope:LFI20:Rad:S}{$-0.95$}
\setsymbol{LFI:slope:LFI21:Rad:S}{$-0.92$}
\setsymbol{LFI:slope:LFI22:Rad:S}{$-1.01$}
\setsymbol{LFI:slope:LFI23:Rad:S}{$-0.95$}
\setsymbol{LFI:slope:LFI24:Rad:S}{$-0.73$}
\setsymbol{LFI:slope:LFI25:Rad:S}{$-1.16$}
\setsymbol{LFI:slope:LFI26:Rad:S}{$-0.79$}
\setsymbol{LFI:slope:LFI27:Rad:S}{$-0.82$}
\setsymbol{LFI:slope:LFI28:Rad:S}{$-0.91$}

\setsymbol{LFI:FWHM:70GHz:units}{13\parcm01}
\setsymbol{LFI:FWHM:44GHz:units}{27\parcm92}
\setsymbol{LFI:FWHM:30GHz:units}{32\parcm65}

\setsymbol{LFI:FWHM:70GHz}{13.01}
\setsymbol{LFI:FWHM:44GHz}{27.92}
\setsymbol{LFI:FWHM:30GHz}{32.65}

\setsymbol{LFI:FWHM:LFI18:units}{13\parcm39}
\setsymbol{LFI:FWHM:LFI19:units}{13\parcm01}
\setsymbol{LFI:FWHM:LFI20:units}{12\parcm75}
\setsymbol{LFI:FWHM:LFI21:units}{12\parcm74}
\setsymbol{LFI:FWHM:LFI22:units}{12\parcm87}
\setsymbol{LFI:FWHM:LFI23:units}{13\parcm27}
\setsymbol{LFI:FWHM:LFI24:units}{22\parcm98}
\setsymbol{LFI:FWHM:LFI25:units}{30\parcm46}
\setsymbol{LFI:FWHM:LFI26:units}{30\parcm31}
\setsymbol{LFI:FWHM:LFI27:units}{32\parcm65}
\setsymbol{LFI:FWHM:LFI28:units}{32\parcm66}

\setsymbol{LFI:FWHM:LFI18}{13.39}
\setsymbol{LFI:FWHM:LFI19}{13.01}
\setsymbol{LFI:FWHM:LFI20}{12.75}
\setsymbol{LFI:FWHM:LFI21}{12.74}
\setsymbol{LFI:FWHM:LFI22}{12.87}
\setsymbol{LFI:FWHM:LFI23}{13.27}
\setsymbol{LFI:FWHM:LFI24}{22.98}
\setsymbol{LFI:FWHM:LFI25}{30.46}
\setsymbol{LFI:FWHM:LFI26}{30.31}
\setsymbol{LFI:FWHM:LFI27}{32.65}
\setsymbol{LFI:FWHM:LFI28}{32.66}

\setsymbol{LFI:FWHM:uncertainty:LFI18:units}{0.170\arcm}
\setsymbol{LFI:FWHM:uncertainty:LFI19:units}{0.174\arcm}
\setsymbol{LFI:FWHM:uncertainty:LFI20:units}{0.170\arcm}
\setsymbol{LFI:FWHM:uncertainty:LFI21:units}{0.156\arcm}
\setsymbol{LFI:FWHM:uncertainty:LFI22:units}{0.164\arcm}
\setsymbol{LFI:FWHM:uncertainty:LFI23:units}{0.171\arcm}
\setsymbol{LFI:FWHM:uncertainty:LFI24:units}{0.652\arcm}
\setsymbol{LFI:FWHM:uncertainty:LFI25:units}{1.075\arcm}
\setsymbol{LFI:FWHM:uncertainty:LFI26:units}{1.131\arcm}
\setsymbol{LFI:FWHM:uncertainty:LFI27:units}{1.266\arcm}
\setsymbol{LFI:FWHM:uncertainty:LFI28:units}{1.287\arcm}

\setsymbol{LFI:FWHM:uncertainty:LFI18}{0.170}
\setsymbol{LFI:FWHM:uncertainty:LFI19}{0.174}
\setsymbol{LFI:FWHM:uncertainty:LFI20}{0.170}
\setsymbol{LFI:FWHM:uncertainty:LFI21}{0.156}
\setsymbol{LFI:FWHM:uncertainty:LFI22}{0.164}
\setsymbol{LFI:FWHM:uncertainty:LFI23}{0.171}
\setsymbol{LFI:FWHM:uncertainty:LFI24}{0.652}
\setsymbol{LFI:FWHM:uncertainty:LFI25}{1.075}
\setsymbol{LFI:FWHM:uncertainty:LFI26}{1.131}
\setsymbol{LFI:FWHM:uncertainty:LFI27}{1.266}
\setsymbol{LFI:FWHM:uncertainty:LFI28}{1.287}

\setsymbol{HFI:center:frequency:100GHz:units}{100\,GHz}
\setsymbol{HFI:center:frequency:143GHz:units}{143\,GHz}
\setsymbol{HFI:center:frequency:217GHz:units}{217\,GHz}
\setsymbol{HFI:center:frequency:353GHz:units}{353\,GHz}
\setsymbol{HFI:center:frequency:545GHz:units}{545\,GHz}
\setsymbol{HFI:center:frequency:857GHz:units}{857\,GHz}

\setsymbol{HFI:center:frequency:100GHz}{100}
\setsymbol{HFI:center:frequency:143GHz}{143}
\setsymbol{HFI:center:frequency:217GHz}{217}
\setsymbol{HFI:center:frequency:353GHz}{353}
\setsymbol{HFI:center:frequency:545GHz}{545}
\setsymbol{HFI:center:frequency:857GHz}{857}

\setsymbol{HFI:Ndetectors:100GHz}{8}
\setsymbol{HFI:Ndetectors:143GHz}{11}
\setsymbol{HFI:Ndetectors:217GHz}{12}
\setsymbol{HFI:Ndetectors:353GHz}{12}
\setsymbol{HFI:Ndetectors:545GHz}{3}
\setsymbol{HFI:Ndetectors:857GHz}{4}

\setsymbol{HFI:FWHM:Maps:100GHz:units}{9\parcm88}
\setsymbol{HFI:FWHM:Maps:143GHz:units}{7\parcm18}
\setsymbol{HFI:FWHM:Maps:217GHz:units}{4\parcm87}
\setsymbol{HFI:FWHM:Maps:353GHz:units}{4\parcm65}
\setsymbol{HFI:FWHM:Maps:545GHz:units}{4\parcm72}
\setsymbol{HFI:FWHM:Maps:857GHz:units}{4\parcm39}
\setsymbol{HFI:FWHM:Maps:100GHz}{9.88}
\setsymbol{HFI:FWHM:Maps:143GHz}{7.18}
\setsymbol{HFI:FWHM:Maps:217GHz}{4.87}
\setsymbol{HFI:FWHM:Maps:353GHz}{4.65}
\setsymbol{HFI:FWHM:Maps:545GHz}{4.72}
\setsymbol{HFI:FWHM:Maps:857GHz}{4.39}

\setsymbol{HFI:beam:ellipticity:Maps:100GHz}{1.15}
\setsymbol{HFI:beam:ellipticity:Maps:143GHz}{1.01}
\setsymbol{HFI:beam:ellipticity:Maps:217GHz}{1.06}
\setsymbol{HFI:beam:ellipticity:Maps:353GHz}{1.05}
\setsymbol{HFI:beam:ellipticity:Maps:545GHz}{1.14}
\setsymbol{HFI:beam:ellipticity:Maps:857GHz}{1.19}

\setsymbol{HFI:FWHM:Mars:100GHz:units}{9\parcm37}
\setsymbol{HFI:FWHM:Mars:143GHz:units}{7\parcm04}
\setsymbol{HFI:FWHM:Mars:217GHz:units}{4\parcm68}
\setsymbol{HFI:FWHM:Mars:353GHz:units}{4\parcm43}
\setsymbol{HFI:FWHM:Mars:545GHz:units}{3\parcm80}
\setsymbol{HFI:FWHM:Mars:857GHz:units}{3\parcm67}

\setsymbol{HFI:FWHM:Mars:100GHz}{9.37}
\setsymbol{HFI:FWHM:Mars:143GHz}{7.04}
\setsymbol{HFI:FWHM:Mars:217GHz}{4.68}
\setsymbol{HFI:FWHM:Mars:353GHz}{4.43}
\setsymbol{HFI:FWHM:Mars:545GHz}{3.80}
\setsymbol{HFI:FWHM:Mars:857GHz}{3.67}

\setsymbol{HFI:beam:ellipticity:Mars:100GHz}{1.18}
\setsymbol{HFI:beam:ellipticity:Mars:143GHz}{1.03}
\setsymbol{HFI:beam:ellipticity:Mars:217GHz}{1.14}
\setsymbol{HFI:beam:ellipticity:Mars:353GHz}{1.09}
\setsymbol{HFI:beam:ellipticity:Mars:545GHz}{1.25}
\setsymbol{HFI:beam:ellipticity:Mars:857GHz}{1.03}

\setsymbol{HFI:CMB:relative:calibration:100GHz}{$\lsim 1\%$}
\setsymbol{HFI:CMB:relative:calibration:143GHz}{$\lsim 1\%$}
\setsymbol{HFI:CMB:relative:calibration:217GHz}{$\lsim 1\%$}
\setsymbol{HFI:CMB:relative:calibration:353GHz}{$\lsim 1\%$}
\setsymbol{HFI:CMB:relative:calibration:545GHz}{}
\setsymbol{HFI:CMB:relative:calibration:857GHz}{}

\setsymbol{HFI:CMB:absolute:calibration:100GHz}{$\lsim 2\%$}
\setsymbol{HFI:CMB:absolute:calibration:143GHz}{$\lsim 2\%$}
\setsymbol{HFI:CMB:absolute:calibration:217GHz}{$\lsim 2\%$}
\setsymbol{HFI:CMB:absolute:calibration:353GHz}{$\lsim 2\%$}
\setsymbol{HFI:CMB:absolute:calibration:545GHz}{}
\setsymbol{HFI:CMB:absolute:calibration:857GHz}{}

\setsymbol{HFI:FIRAS:gain:calibration:accuracy:statistical:100GHz}{}
\setsymbol{HFI:FIRAS:gain:calibration:accuracy:statistical:143GHz}{}
\setsymbol{HFI:FIRAS:gain:calibration:accuracy:statistical:217GHz}{}
\setsymbol{HFI:FIRAS:gain:calibration:accuracy:statistical:353GHz}{2.5\%}
\setsymbol{HFI:FIRAS:gain:calibration:accuracy:statistical:545GHz}{1\%}
\setsymbol{HFI:FIRAS:gain:calibration:accuracy:statistical:857GHz}{0.5\%}

\setsymbol{HFI:FIRAS:gain:calibration:accuracy:systematic:100GHz}{}
\setsymbol{HFI:FIRAS:gain:calibration:accuracy:systematic:143GHz}{}
\setsymbol{HFI:FIRAS:gain:calibration:accuracy:systematic:217GHz}{}
\setsymbol{HFI:FIRAS:gain:calibration:accuracy:systematic:353GHz}{}
\setsymbol{HFI:FIRAS:gain:calibration:accuracy:systematic:545GHz}{7\%}
\setsymbol{HFI:FIRAS:gain:calibration:accuracy:systematic:857GHz}{7\%}

\setsymbol{HFI:FIRAS:zero:point:accuracy:100GHz:units}{0.8\MJysr}
\setsymbol{HFI:FIRAS:zero:point:accuracy:143GHz:units}{}
\setsymbol{HFI:FIRAS:zero:point:accuracy:217GHz:units}{}
\setsymbol{HFI:FIRAS:zero:point:accuracy:353GHz:units}{1.4\MJysr}
\setsymbol{HFI:FIRAS:zero:point:accuracy:545GHz:units}{2.2\MJysr}
\setsymbol{HFI:FIRAS:zero:point:accuracy:857GHz:units}{1.7\MJysr}

\setsymbol{HFI:FIRAS:zero:point:accuracy:100GHz}{0.8}
\setsymbol{HFI:FIRAS:zero:point:accuracy:143GHz}{}
\setsymbol{HFI:FIRAS:zero:point:accuracy:217GHz}{}
\setsymbol{HFI:FIRAS:zero:point:accuracy:353GHz}{1.4}
\setsymbol{HFI:FIRAS:zero:point:accuracy:545GHz}{2.2}
\setsymbol{HFI:FIRAS:zero:point:accuracy:857GHz}{1.7}

\setsymbol{HFI:unit:conversion:100GHz:units}{0.2415\MJysrmK}
\setsymbol{HFI:unit:conversion:143GHz:units}{0.3694\MJysrmK}
\setsymbol{HFI:unit:conversion:217GHz:units}{0.4811\MJysrmK}
\setsymbol{HFI:unit:conversion:353GHz:units}{0.2883\MJysrmK}
\setsymbol{HFI:unit:conversion:545GHz:units}{0.05826\MJysrmK}
\setsymbol{HFI:unit:conversion:857GHz:units}{0.002238\MJysrmK}

\setsymbol{HFI:unit:conversion:100GHz}{0.2415}
\setsymbol{HFI:unit:conversion:143GHz}{0.3694}
\setsymbol{HFI:unit:conversion:217GHz}{0.4811}
\setsymbol{HFI:unit:conversion:353GHz}{0.2883}
\setsymbol{HFI:unit:conversion:545GHz}{0.05826}
\setsymbol{HFI:unit:conversion:857GHz}{0.002238}

\setsymbol{HFI:colour:correction:alpha=-2:V1.01:100GHz}{0.9893}
\setsymbol{HFI:colour:correction:alpha=-2:V1.01:143GHz}{0.9759}
\setsymbol{HFI:colour:correction:alpha=-2:V1.01:217GHz}{1.0007}
\setsymbol{HFI:colour:correction:alpha=-2:V1.01:353GHz}{1.0028}
\setsymbol{HFI:colour:correction:alpha=-2:V1.01:545GHz}{1.0019}
\setsymbol{HFI:colour:correction:alpha=-2:V1.01:857GHz}{0.9889}

\setsymbol{HFI:colour:correction:alpha=0:V1.01:100GHz}{1.0008}
\setsymbol{HFI:colour:correction:alpha=0:V1.01:143GHz}{1.0148}
\setsymbol{HFI:colour:correction:alpha=0:V1.01:217GHz}{0.9909}
\setsymbol{HFI:colour:correction:alpha=0:V1.01:353GHz}{0.9888}
\setsymbol{HFI:colour:correction:alpha=0:V1.01:545GHz}{0.9878}
\setsymbol{HFI:colour:correction:alpha=0:V1.01:857GHz}{1.0014}

%% file: 16455_b.tex
\author{\small
Planck Collaboration:
A.~Abergel\inst{47}
\and
P.~A.~R.~Ade\inst{72}
\and
N.~Aghanim\inst{47}
\and
M.~Arnaud\inst{58}
\and
M.~Ashdown\inst{56, 4}
\and
J.~Aumont\inst{47}
\and
C.~Baccigalupi\inst{70}
\and
A.~Balbi\inst{28}
\and
A.~J.~Banday\inst{76, 7, 63}
\and
R.~B.~Barreiro\inst{53}
% ...
}
\institute{\small
Aalto University Mets\"{a}hovi Radio Observatory, Mets\"{a}hovintie 114, FIN-02540 Kylm\"{a}l\"{a}, Finland\\
\and
Agenzia Spaziale Italiana Science Data Center, c/o ESRIN, via Galileo Galilei, Frascati, Italy\\
\and
Astroparticule et Cosmologie, CNRS (UMR7164), Universit\'{e} Denis Diderot Paris 7, B\^{a}timent Condorcet, 10 rue A. Domon et L\'{e}onie Duquet, Paris, France\\
\and
Astrophysics Group, Cavendish Laboratory, University of Cambridge, J J Thomson Avenue, Cambridge CB3 0HE, U.K.\\
\and
Atacama Large Millimeter/submillimeter Array, ALMA Santiago Central Offices, Alonso de Cordova 3107, Vitacura, Casilla 763 0355, Santiago, Chile\\
\and
CITA, University of Toronto, 60 St. George St., Toronto, ON M5S 3H8, Canada\\
\and
CNRS, IRAP, 9 Av. colonel Roche, BP 44346, F-31028 Toulouse cedex 4, France\\
\and
California Institute of Technology, Pasadena, California, U.S.A.\\
\and
Centre of Mathematics for Applications, University of Oslo, Blindern, Oslo, Norway\\
\and
DAMTP, University of Cambridge, Centre for Mathematical Sciences, Wilberforce Road, Cambridge CB3 0WA, U.K.\\
\and
DSM/Irfu/SPP, CEA-Saclay, F-91191 Gif-sur-Yvette Cedex, France\\
\and
DTU Space, National Space Institute, Juliane Mariesvej 30, Copenhagen, Denmark\\
\and
Departamento de F\'{\i}sica, Universidad de Oviedo, Avda. Calvo Sotelo s/n, Oviedo, Spain\\
\and
Department of Astronomy and Astrophysics, University of Toronto, 50 Saint George Street, Toronto, Ontario, Canada\\
\and
Department of Physics \& Astronomy, University of British Columbia, 6224 Agricultural Road, Vancouver, British Columbia, Canada\\
\and
Department of Physics, Gustaf H\"{a}llstr\"{o}min katu 2a, University of Helsinki, Helsinki, Finland\\
\and
Department of Physics, Princeton University, Princeton, New Jersey, U.S.A.\\
\and
Department of Physics, Purdue University, 525 Northwestern Avenue, West Lafayette, Indiana, U.S.A.\\
\and
Department of Physics, University of California, Berkeley, California, U.S.A.\\
\and
Department of Physics, University of California, One Shields Avenue, Davis, California, U.S.A.\\
\and
Department of Physics, University of California, Santa Barbara, California, U.S.A.\\
\and
Department of Physics, University of Illinois at Urbana-Champaign, 1110 West Green Street, Urbana, Illinois, U.S.A.\\
\and
Dipartimento di Fisica G. Galilei, Universit\`{a} degli Studi di Padova, via Marzolo 8, 35131 Padova, Italy\\
\and
Dipartimento di Fisica, Universit\`{a} La Sapienza, P. le A. Moro 2, Roma, Italy\\
\and
Dipartimento di Fisica, Universit\`{a} degli Studi di Milano, Via Celoria, 16, Milano, Italy\\
\and
Dipartimento di Fisica, Universit\`{a} degli Studi di Trieste, via A. Valerio 2, Trieste, Italy\\
\and
Dipartimento di Fisica, Universit\`{a} di Ferrara, Via Saragat 1, 44122 Ferrara, Italy\\
\and
Dipartimento di Fisica, Universit\`{a} di Roma Tor Vergata, Via della Ricerca Scientifica, 1, Roma, Italy\\
\and
Discovery Center, Niels Bohr Institute, Blegdamsvej 17, Copenhagen, Denmark\\
\and
Dpto. Astrof\'{i}sica, Universidad de La Laguna (ULL), E-38206 La Laguna, Tenerife, Spain\\
\and
European Southern Observatory, ESO Vitacura, Alonso de Cordova 3107, Vitacura, Casilla 19001, Santiago, Chile\\
\and
European Space Agency, ESAC, Planck Science Office, Camino bajo del Castillo, s/n, Urbanizaci\'{o}n Villafranca del Castillo, Villanueva de la Ca\~{n}ada, Madrid, Spain\\
\and
European Space Agency, ESTEC, Keplerlaan 1, 2201 AZ Noordwijk, The Netherlands\\
\and
Harvard-Smithsonian Center for Astrophysics, 60 Garden Street, Cambridge, MA 02138, U.S.A.\\
\and
Helsinki Institute of Physics, Gustaf H\"{a}llstr\"{o}min katu 2, University of Helsinki, Helsinki, Finland\\
\and
INAF - Osservatorio Astrofisico di Catania, Via S. Sofia 78, Catania, Italy\\
\and
INAF - Osservatorio Astronomico di Padova, Vicolo dell'Osservatorio 5, Padova, Italy\\
\and
INAF - Osservatorio Astronomico di Roma, via di Frascati 33, Monte Porzio Catone, Italy\\
\and
INAF - Osservatorio Astronomico di Trieste, Via G.B. Tiepolo 11, Trieste, Italy\\
\and
INAF/IASF Bologna, Via Gobetti 101, Bologna, Italy\\
\and
INAF/IASF Milano, Via E. Bassini 15, Milano, Italy\\
\and
INRIA, Laboratoire de Recherche en Informatique, Universit\'{e} Paris-Sud 11, B\^{a}timent 490, 91405 Orsay Cedex, France\\
\and
IPAG: Institut de Plan\'{e}tologie et d'Astrophysique de Grenoble, Universit\'{e} Joseph Fourier, Grenoble 1 / CNRS-INSU, UMR 5274, Grenoble, F-38041, France\\
\and
Imperial College London, Astrophysics group, Blackett Laboratory, Prince Consort Road, London, SW7 2AZ, U.K.\\
\and
Infrared Processing and Analysis Center, California Institute of Technology, Pasadena, CA 91125, U.S.A.\\
\and
Institut N\'{e}el, CNRS, Universit\'{e} Joseph Fourier Grenoble I, 25 rue des Martyrs, Grenoble, France\\
\and
Institut d'Astrophysique Spatiale, CNRS (UMR8617) Universit\'{e} Paris-Sud 11, B\^{a}timent 121, Orsay, France\\
\and
Institut d'Astrophysique de Paris, CNRS UMR7095, Universit\'{e} Pierre \& Marie Curie, 98 bis boulevard Arago, Paris, France\\
\and
Institute of Astronomy and Astrophysics, Academia Sinica, Taipei, Taiwan\\
\and
Institute of Astronomy, University of Cambridge, Madingley Road, Cambridge CB3 0HA, U.K.\\
\and
Institute of Theoretical Astrophysics, University of Oslo, Blindern, Oslo, Norway\\
\and
Instituto de Astrof\'{\i}sica de Canarias, C/V\'{\i}a L\'{a}ctea s/n, La Laguna, Tenerife, Spain\\
\and
Instituto de F\'{\i}sica de Cantabria (CSIC-Universidad de Cantabria), Avda. de los Castros s/n, Santander, Spain\\
\and
Jet Propulsion Laboratory, California Institute of Technology, 4800 Oak Grove Drive, Pasadena, California, U.S.A.\\
\and
Jodrell Bank Centre for Astrophysics, Alan Turing Building, School of Physics and Astronomy, The University of Manchester, Oxford Road, Manchester, M13 9PL, U.K.\\
\and
Kavli Institute for Cosmology Cambridge, Madingley Road, Cambridge, CB3 0HA, U.K.\\
\and
LERMA, CNRS, Observatoire de Paris, 61 Avenue de l'Observatoire, Paris, France\\
\and
Laboratoire AIM, IRFU/Service d'Astrophysique - CEA/DSM - CNRS - Universit\'{e} Paris Diderot, B\^{a}t. 709, CEA-Saclay, F-91191 Gif-sur-Yvette Cedex, France\\
\and
Laboratoire Traitement et Communication de l'Information, CNRS (UMR 5141) and T\'{e}l\'{e}com ParisTech, 46 rue Barrault F-75634 Paris Cedex 13, France\\
\and
Laboratoire de Physique Subatomique et de Cosmologie, CNRS/IN2P3, Universit\'{e} Joseph Fourier Grenoble I, Institut National Polytechnique de Grenoble, 53 rue des Martyrs, 38026 Grenoble cedex, France\\
\and
Laboratoire de l'Acc\'{e}l\'{e}rateur Lin\'{e}aire, Universit\'{e} Paris-Sud 11, CNRS/IN2P3, Orsay, France\\
\and
Lawrence Berkeley National Laboratory, Berkeley, California, U.S.A.\\
\and
Max-Planck-Institut f\"{u}r Astrophysik, Karl-Schwarzschild-Str. 1, 85741 Garching, Germany\\
\and
Max-Planck-Institut f\"{u}r Radioastronomie, Auf dem H\"{u}gel 69, 53121 Bonn, Germany\\
\and
MilliLab, VTT Technical Research Centre of Finland, Tietotie 3, Espoo, Finland\\
\and
National University of Ireland, Department of Experimental Physics, Maynooth, Co. Kildare, Ireland\\
\and
Niels Bohr Institute, Blegdamsvej 17, Copenhagen, Denmark\\
\and
Observational Cosmology, Mail Stop 367-17, California Institute of Technology, Pasadena, CA, 91125, U.S.A.\\
\and
Optical Science Laboratory, University College London, Gower Street, London, U.K.\\
\and
SISSA, Astrophysics Sector, via Bonomea 265, 34136, Trieste, Italy\\
\and
SUPA, Institute for Astronomy, University of Edinburgh, Royal Observatory, Blackford Hill, Edinburgh EH9 3HJ, U.K.\\
\and
School of Physics and Astronomy, Cardiff University, Queens Buildings, The Parade, Cardiff, CF24 3AA, U.K.\\
\and
Space Sciences Laboratory, University of California, Berkeley, California, U.S.A.\\
\and
Spitzer Science Center, 1200 E. California Blvd., Pasadena, California, U.S.A.\\
\and
Stanford University, Dept of Physics, Varian Physics Bldg, 382 Via Pueblo Mall, Stanford, California, U.S.A.\\
\and
Universit\'{e} de Toulouse, UPS-OMP, IRAP, F-31028 Toulouse cedex 4, France\\
% ...
}